\documentclass[journal,twoside,web]{./Common/ieeecolor}
\usepackage{jsen}
\usepackage{cite}
\usepackage{amsmath,amssymb,amsfonts}
\usepackage{algorithmic}
\usepackage{graphicx}
\usepackage{lipsum} 
\usepackage{textcomp}
\usepackage{wrapfig}
\usepackage{dblfloatfix}    % To enable figures at the bottom of page, EG

\usepackage[hidelinks]{hyperref}
\def\BibTeX{{\rm B\kern-.05em{\sc i\kern-.025em b}\kern-.08em
    T\kern-.1667em\lower.7ex\hbox{E}\kern-.125emX}}
\definecolor{abstractbg}{rgb}{0.89804,0.94510,0.83137}
\usepackage{xspace}
\xspaceaddexceptions{-}
\usepackage{glossaries}
\usepackage{color}
\usepackage{graphicx}        % standard LaTeX graphics tool
\graphicspath{{./figures/}}
\usepackage{subfigure}
\usepackage{cite}

\let\OLDthebibliography\thebibliography
\renewcommand\thebibliography[1]{
	\OLDthebibliography{#1}
	\setlength{\parskip}{0pt}
	\setlength{\itemsep}{0pt}
}

\usepackage{siunitx}
\usepackage{amsmath,amssymb,latexsym}
\usepackage{amsfonts}
\usepackage{upgreek}
\usepackage{mathtools}
\usepackage{nicefrac}

\usepackage{booktabs}
\usepackage{multirow}
\usepackage[table,xcdraw]{xcolor}
\usepackage[flushleft]{threeparttable}

\newacronym{ASIen}{ASI}{Italian Space Agency}
    
\newacronym{ASIit}{ASI}{Agenzia Spaziale Italiana}
    
\newacronym{UNICALen}{UNICAL}{University of Calabria}
    
\newacronym{UNICALit}{UNICAL}{Università della Calabria}
    
\newacronym{DIMESen}{DIMES}{Department of Computer Engineering, Modeling, Electronics and Systems}
    
\newacronym{DIMESit}{DIMES}{Dipartimento di Ingegneria Informatica, Modellistica, Elettronica e Sistemistica}
    
\newacronym{USFQes}{USFQ}{Universidad San Francisco de Quito}
         
\newacronym{BIUen}{BIU}{Bar-Ilan University}
     
\newacronym{tsmc}{TSMC}{Taiwan Semiconductor Manufacturing Company}
     
\newacronym{beol}{BEOL}{back-end-of-line}

\newacronym{feol}{FEOL}{front-end-of-line}

\newacronym{wordl}{WL}{wordline}
    
\newacronym{bitl}{BL}{bitline}
    
\newacronym{sourcel}{SL}{sourceline}
    
\newacronym[longplural={nanowires}]{nw}{NW}{nanowire}

\newacronym{ptm}{PTM}{predictive technology model}

\newacronym{pdk}{PDK}{process design kit}

\newacronym[longplural={systems-on-chip}]{soc}{SoC}{system-on-chip}

\newacronym[longplural={integrated circuits}]{ic}{IC}{integrated circuit}

\newacronym{ai}{AI}{artificial-intelligence}

\newacronym{iot}{IoT}{internet-of-things}

\newacronym{mc}{MC}{Monte Carlo}
     
\newacronym{cvs}{CVS}{conventional voltage sensing}

\newacronym{epi}{EPI}{energy per instruction}

\newacronym{ips}{IPS}{instructions per second}

\newacronym{mep}{MEP}{minimum energy point}

\newacronym{lrs}{LRS}{low resistance state}

\newacronym{hrs}{HRS}{high resistance state}

\newacronym{mim}{MIM}{metal-insulator-metal}

\newacronym[longplural={phase-change memories}]{pcm}{PCM}{phase-change memory}

\newacronym[longplural={resistive RAMs}]{rram}{RRAM}{resistive RAM}

\newacronym[longplural={spin-transfer torque magnetic RAMs}]{sttmram}{STT-MRAM}{spin-transfer torque magnetic RAM}

\newacronym{euv}{EUV}{extreme ultra-violet}
     
\newacronym[longplural={Gain-Cell embedded DRAMs}]{gcedram}{GC-eDRAM}{Gain-Cell embedded DRAM}

\newacronym{sixt}{6T}{6-transistor}
    
\newacronym{eflash}{eFlash}{embedded Flash}

\newacronym[longplural={multi-level cells}]{mlc}{MLC}{multi-level cell}

\newacronym[longplural={Storage Class Memories}]{scm}{SCM}{Storage Class Memory}

\newacronym{ddr}{DDR}{dual-data rate}

\newacronym[longplural={graphic processing units}]{gpu}{GPU}{graphic processing unit}

\newacronym[longplural={central processing units}]{cpu}{CPU}{central processing unit}

\newacronym{sata}{SATA}{Serial Advanced Technology Attachment}
    
\newacronym{nvme}{NVMe}{Non-Volatile Memory Express}
    
\newacronym{nvm}{NVM}{NVM}
        
\newacronym{pcie}{PCIe}{Peripheral Component Interconnect Express}
     
\newacronym[longplural={hard-Disk drives}]{hdd}{HDD}{hard-Disk drive}

\newacronym[longplural={solid-State drives}]{ssd}{SSD}{solid-State drive}

\newacronym[longplural={high-bandwidth memories}]{hbm}{HBM}{high-bandwidth memory}

\newacronym[longplural={dual-inline memory modules}]{dimm}{DIMM}{dual-inline memory module}

\newacronym[longplural={static random-access memories}]{sram}{SRAM}{static random-access memory}

\newacronym[longplural={embedded DRAMs}]{edram}{eDRAM}{embedded DRAM}

\newacronym[longplural={dynamic random-access memories}]{dram}{DRAM}{dynamic random-access memory}

\newacronym[longplural={six-transistor static random access memories}]{sixtsram}{6T-SRAM}{six-transistor static random access memory}

\newacronym[longplural={magnetic random-access memories}]{mram}{MRAM}{magnetic random-access memory}

\newacronym{bc}{BC}{bitcell}

\newacronym{bl}{BL}{bitline}

\newacronym{sln}{SL}{sourceline}

\newacronym{wl}{WL}{wordline}

\newacronym{nc}{NC}{number of cycles at endurance failure}
     
\newacronym{llgs}{LLGS}{Landau-Lifshitz-Gilbert-Slonczewski}
     
\newacronym{stt}{STT}{spin-transfer torque}

\newacronym{pma}{PMA}{perpendicular magnetic anisotropy}

\newacronym{ima}{IMA}{in-plane magnetic anisotropy}

\newacronym{mtj}{MTJ}{magnetic tunnel junction}

\newacronym{smtj}{SMTJ}{single-barrier MTJ}

\newacronym{dmtj}{DMTJ}{double-barrier MTJ}

\newacronym{mr}{MR}{magnetoresistance}

\newacronym{tmr}{TMR}{tunnel magnetoresistance}

\newacronym{gmr}{GMR}{giant magnetoresistance}

\newacronym{wer}{WER}{write error rate}

\newacronym{rdr}{RDR}{read disturbance rate}

\newacronym{rfr}{RFR}{retention failure rate}

\newacronym{rer}{RER}{read error rate}

\newacronym{fl}{FL}{free layer}

\newacronym[longplural={reference layers}]{rl}{RL}{reference layer}

\newacronym{p}{P}{parallel}

\newacronym{ap}{AP}{antiparallel}

\newacronym{fm}{FM}{ferromagnetic}

\newcommand{\uA}{\,\si{\micro\ampere}\xspace}

\newcommand{\V}{\,\si{\volt}\xspace}
\newcommand{\mV}{\,\si{\milli\volt}\xspace}

\newcommand{\fF}{\,\si{\femto\farad}\xspace}

\newcommand{\nJ}{\,\si{\nano\joule}\xspace}

\newcommand{\mmsquared}{\,\si{\milli\meter\squared}\xspace}

\newcommand{\nm}{\,\si{\nano\meter}\xspace}

\newcommand{\umsquared}{\,\si{\micro\meter\squared}\xspace}

\newcommand{\bit}{\,\si{\bit}\xspace}

\newcommand{\kHz}{\,\si{\kilo\hertz}\xspace}
\newcommand{\MHz}{\,\si{\mega\hertz}\xspace}

\newcommand{\uW}{\,\si{\micro\watt}\xspace}

\newcommand{\C}{\,\si{\celsius}\xspace}

\renewcommand{\eqref}[1]{(\ref{#1})}

\newcommand{\figref}[1]{\mbox{Fig.~\ref{#1}}}

\usepackage{subfiles} % Best loaded last in the preamble

\usepackage{etoolbox}
\makeatletter
\@ifundefined{color@begingroup}%
  {\let\color@begingroup\relax
   \let\color@endgroup\relax}{}%
\def\fix@ieeecolor@hbox#1{%
  \hbox{\color@begingroup#1\color@endgroup}}
\patchcmd\@makecaption{\hbox}{\fix@ieeecolor@hbox}{}{\FAILED}
\patchcmd\@makecaption{\hbox}{\fix@ieeecolor@hbox}{}{\FAILED}
\begin{document}

\title{A 0.6\V--1.8\V Compact Temperature Sensor with 0.24\C Resolution, $\pm$1.4\C Inaccuracy and 1.06\nJ per Conversion}
% %========ORCID of the authors
\newcommand{\orcidBZ}{\href{https://orcid.org/0000-0002-1301-3447}{\includegraphics[scale=0.01]{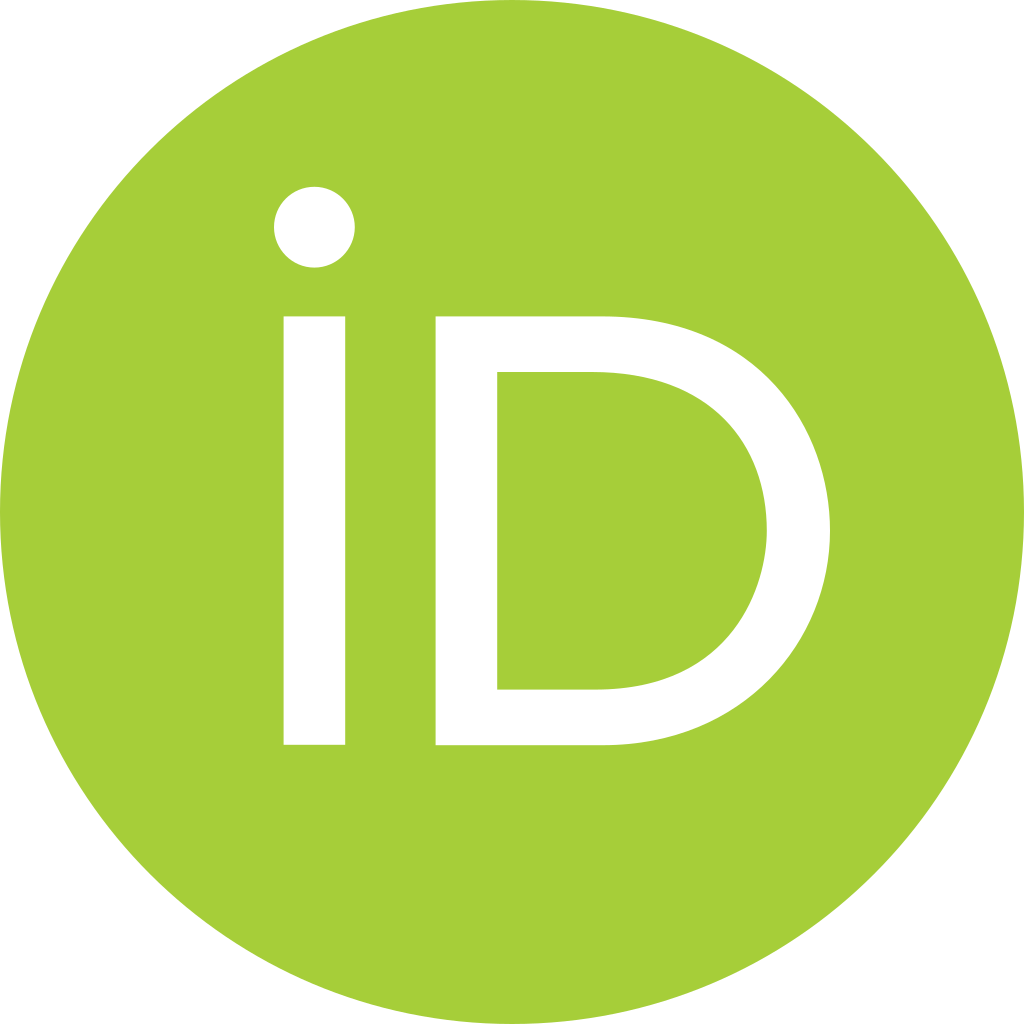}}}
\newcommand{\orcidEG}{\href{https://orcid.org/0000-0002-5862-2246}{\includegraphics[scale=0.01]{ORCID_icon.png}}}
\newcommand{\orcidSeS}{\href{https://orcid.org/0000-0002-6984-1137}{\includegraphics[scale=0.01]{ORCID_icon.png}}}
\newcommand{\orcidML}{\href{https://orcid.org/0000-0002-6480-9218}{\includegraphics[scale=0.01]{ORCID_icon.png}}}
\newcommand{\orcidGI}{\href{https://orcid.org/0000-0003-3375-1647}{\includegraphics[scale=0.01]{ORCID_icon.png}}}
% %=============================

\author{Benjamin~Zambrano$^{\orcidBZ}$, Esteban~Garzón$^{\orcidEG}$, \IEEEmembership{Graduate Student Member, IEEE}, Sebastiano~Strangio$^{\orcidSeS}$, \IEEEmembership{Member, IEEE}, Giuseppe~Iannaccone$^{\orcidGI}$, \IEEEmembership{Fellow, IEEE}, and Marco~Lanuzza$^{\orcidML}$, \IEEEmembership{Senior Member, IEEE} 
\thanks{Manuscript received February. XXX, 2022. This work has been partially supported by the Ministero dell’Istruzione, dell’Università e
della Ricerca (MIUR) CrossLab Departments of Excellence Grant, and in part by the ECSEL Joint Undertaking (JU) under grant agreement No 876362. The JU receives support from the European Union’s Horizon 2020 research and innovation programme and Finland, Austria, Belgium, Czechia, Germany, Italy, Latvia, Netherlands, Poland, Switzerland.}
\thanks{B. Zambrano, E. Garzón, and M. Lanuzza are with the Dipartimento di Ingegneria Informatica, Modellistica, Elettronica e Sistemistica (DIMES),
        Università della Calabria, 
        Rende 87036, Italy (e-mail:
        \{benjamin.zambrano, esteban.garzon, marco.lanuzza\}@unical.it.}
\thanks{S. Strangio and G. Iannaccone are with the Dipartimento di Ingegneria dell'Informazione, Università di Pisa, Via G. Caruso 16, 56122, Pisa, Italy (e-mail: \{sebastiano.strangio, giuseppe.iannaccone\}@unipi.it).}
\thanks{\copyright~2022 IEEE. Personal use of this material is permitted. Permission from IEEE must be obtained for all other uses, in any current or future
media, including reprinting/republishing this material for advertising or promotional purposes, creating new collective works, for resale or
redistribution to servers or lists, or reuse of any copyrighted component of this work in other works. DOI: 10.1109/JSEN.2022.3171106}
}

\IEEEtitleabstractindextext{%
\fcolorbox{abstractbg}{abstractbg}{%
\begin{minipage}{\textwidth}%
\begin{wrapfigure}[12]{r}{3in}%
\includegraphics[width=3in]{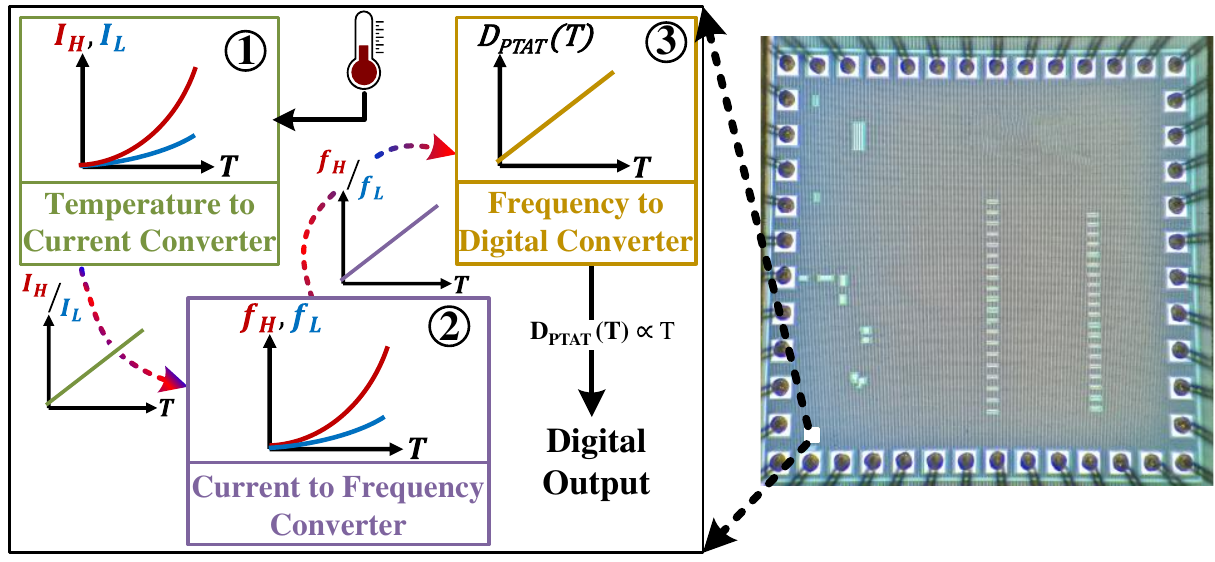}%
\end{wrapfigure}%
\begin{abstract}
This paper presents a fully-integrated CMOS temperature sensor for densely-distributed thermal monitoring in systems on chip supporting dynamic voltage and frequency scaling. The sensor front-end exploits a sub-threshold PMOS-based circuit to convert the local temperature into two biasing currents. These are then used to define two oscillation frequencies, whose ratio is proportional to absolute-temperature. Finally, the sensor back-end translates such frequency ratio into the digital temperature code. Thanks to its low-complexity architecture, the proposed design achieves a very compact footprint along with low-power consumption and high accuracy in a wide temperature range. Moreover, thanks to a simple embedded line regulation mechanism, our sensor supports voltage-scalability. The design was prototyped in a 180\nm CMOS technology with a 0\C $-$ 100\C temperature detection range, a very wide supply voltage operating range from 0.6\V up to 1.8\V and very small silicon area occupation of just 0.021\mmsquared. Experimental measurements performed on 20 test chips have shown very competitive figures of merit, including a resolution of 0.24\C, an inaccuracy of $\pm$1.4\C, a sampling rate of about 1.5\kHz and an energy per conversion of 1.06\nJ at 30\C.
\end{abstract}

\end{minipage}}}

\maketitle
\section{Introduction} \label{sec:introduction}
\IEEEPARstart{T}{emperature} sensors are crucial to allow dynamic thermal management (DTM) in complex Systems on Chips (SoCs)~\cite{li2020accurate, long2008thermal, li2017heuristic, nowroz2010thermal}. Multiple sensors are distributed across the die to identify potential hot/cold-spots, and the collected
temperature information is used to keep the operation of the chip within the target thermal condition, thus ensuring both performance and reliability~\cite{long2008thermal, li2017heuristic, nowroz2010thermal}. The temperature safety can be maintained while accommodating dynamic voltage and frequency scaling (DVFS), powering on/off different portions of the system and/or implementing other adaptive thermal adjustments like fan speed regulation~\cite{li2020accurate, gonzalez20173, Shor2013miniaturized}.

  \begin{figure*}[h!] % [b]-> bottom, [t]->top, [H]->Here! ([h!] should do a better job),     {figure*}->float
     \centering
     \includegraphics[width=0.90\textwidth]{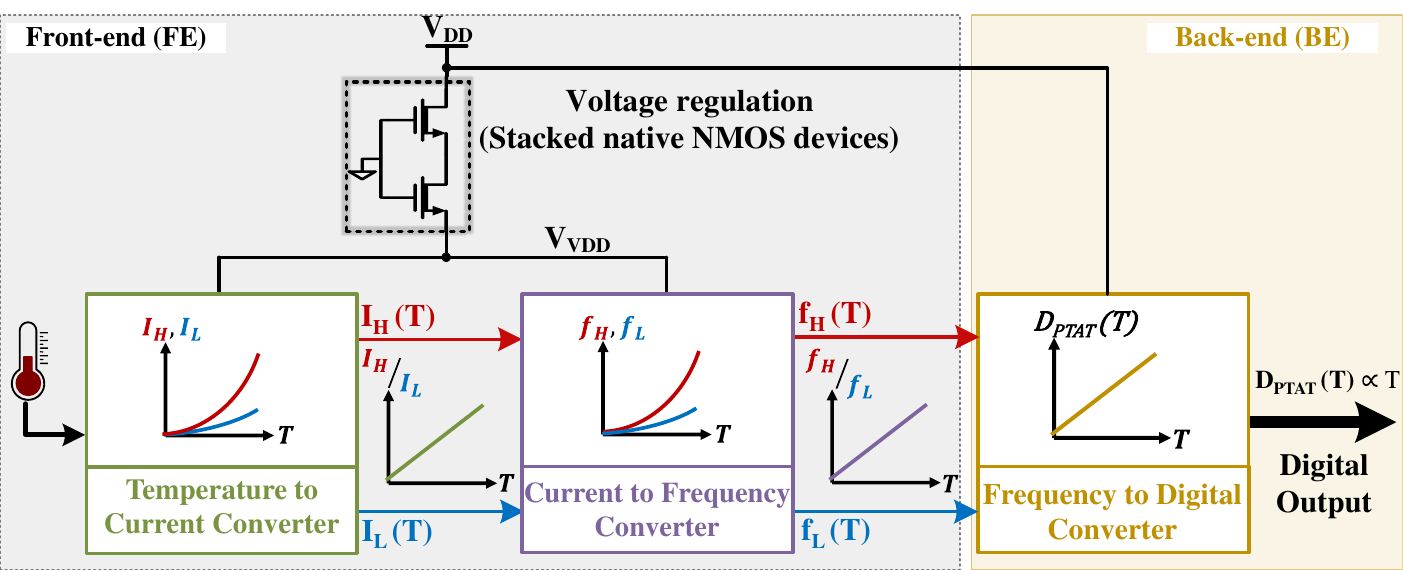}\vspace{-2mm}
     \caption{Block diagram of the temperature sensor.}
     \label{fig:FIG1}
     \vspace{-6mm}
   \end{figure*}

Some proposed temperature sensors~\cite{yang2015compact, vinshtok2020ultra, anand2016vco_JSSC16, wang2019763_JSSC19_763p,li2020a+_TCASI21, yang20179, someya201911, zambrano20210, huang2017energy, an2014energy, Jalalifar2016, azam2021ultra, Tang2018, Hung2016} exhibit the required sensing accuracy for proper DTM. A typical requirement for multi-core systems is a modest absolute inaccuracy of 8\C and a more constrained relative inaccuracy of 3\C~\cite{vinshtok2020ultra,li2009 }. Additional requirements are imposed by recent technology trends, such as multi-processor chips,  3-D integration and multi-supply voltage architectures~\cite{yang2015compact}. Particularly:
\begin{enumerate}
    \item Small size is a highly desired feature to enable very dense thermal monitoring, as required in  state-of-the-art VLSI designs. In fact, due to the ever-increasing level of integration of modern SoCs, the number of possible hot/cold-spots is rapidly growing with the consequent growth in the number of required temperature sensors for effective DTM (e.g., more than 60 in the POWER9 systems~\cite{gonzalez20173}). Moreover, a compact footprint is further important to ensure flexibility in the design process, since the optimal sensor locations (e.g., very close to the possible hot/cold-spots) often are identified only in the later stages of the design phase and,  typically they are in very dense areas of the chip~\cite{vinshtok2020ultra}.
    \item Robustness is essential: A temperature sensor needs to maintain its sensing accuracy under voltage and process variations. In fact, overestimating the temperature can cause unnecessary performance throttling, while underestimating can lead to reliability issues.  
    \item Broad supply voltage scalability is a further requirement~\cite{yang2015compact}. Today’s SoCs often implement DVFS techniques to tune performance while managing power consumption, especially for digital portions of the system. 
    The supply voltage can be dynamically modulated down to near threshold levels when reduced performance allows saving energy. It is therefore desirable that temperature sensors support  supply voltage scalability in order to share the same power grids with the digital circuits. Unfortunately, some of the previously proposed temperature sensors turned out to be not enough voltage scalable~\cite{anand2016vco_JSSC16, li2020a+_TCASI21,zambrano20210, huang2017energy, an2014energy, Tang2018}, and/or do not support sub-1\V operation~\cite{vinshtok2020ultra, Jalalifar2016, Hwang213, huang2017energy, an2014energy, Tang2018}.
\end{enumerate}

This paper introduces a small-area fully-integrated CMOS temperature sensor, suitable for aggressive circuit placement (i.e., very close to target hot/cold-spots) in SoCs. A low-complexity PMOS-based sensing circuit converts the local temperature into two sub-threshold biasing currents. These are then used to define two oscillation frequencies, whose ratio increases linearly as the temperature rises.
Such frequency ratio is then translated into a digital output code, through a digital back-end, based on binary counters. Thanks to a simple embedded line regulation mechanism, the proposed design can operate in a wide power supply range, thus resulting to be appealing for those systems which support multi-supply voltages and/or DVFS. When fabricated in 180\,nm CMOS for the target \mbox{0\C\,--\,100\C} temperature range, our design exhibits a compact footprint of about \mbox{0.02\mmsquared} and a supply voltage operating range from 0.6\,V to 1.8\,V. Moreover, it consumes less than \mbox{1.6\uW} ($V_{DD}$=0.6\,V and $Temp$=\mbox{25\C}), with an energy per conversion of 1.05\nJ. The above results were achieved with an inaccuracy constrained within $\pm$\mbox{1.4\C}, and a resolution of \mbox{0.24\C}.
   
The remainder of the paper is organized as follows. 
Section II details the temperature sensor design.
Section III presents measurement results over 20 test samples. Section IV compares our proposal with some recent CMOS-based competitors proposed in the literature. Finally, Section V concludes the work.

\section{Proposed Temperature Sensor}\label{sec:TSens}

Inherited from the temperature-to-digital conversion methodology exploited in~\cite{zambrano20210, someya201911}, the sensor relies on three main processing blocks, as shown in \figref{fig:FIG1}. 1) The temperature-to-current converter (TCC) represents the sensing element of the proposed circuit. It generates two sub-threshold currents, $I_H$ and $I_L$ (with $I_H > I_L$), whose ratio $I_H/I_L$ has an increasing linear dependence with the temperature. 2) The current-to-frequency converter (CFC) consists of two independent ring oscillators controlled by mirrored $I_H$ and $I_L$ currents. The ratio of the two oscillation frequencies (i.e., $f_H$/$f_L$) is maintained linear with temperature. 3) Finally, the frequency-to-digital converter (FDC) generates the digital temperature code, based on the $f_H$/$f_L$ ratio.

The voltage scalability of the sensor derives from two stacked native (i.e., zero threshold voltage transistors) NMOS devices~\cite{yang20179}, which provide line regulation for the supply-sensitive TCC and CFC blocks. Indeed, these are powered in the sub-threshold region by an almost stable virtual $V_{DD}$ ($V_{VDD}$), regardless of the actual $V_{DD}$ and temperature. In our 180\nm implementation, the $V_{VDD}$ is about 440\mV across very large temperature (0\C -- 100\C) and power supply (0.6\V -- 1.8\V) variations. Differently, the FDC circuit, being based on not-critical digital circuitry (i.e., intrinsically more robust to temperature and voltage variations) is powered directly by the $V_{DD}$.

\subsection{Temperature-to-Current Converter}\label{sec:TCC}

\begin{figure}[!t] % [b]-> bottom, [t]->top, [H]->Here! ([h!] should do a better job),     {figure*}->float
     \centering
     \includegraphics[width=0.49\textwidth]{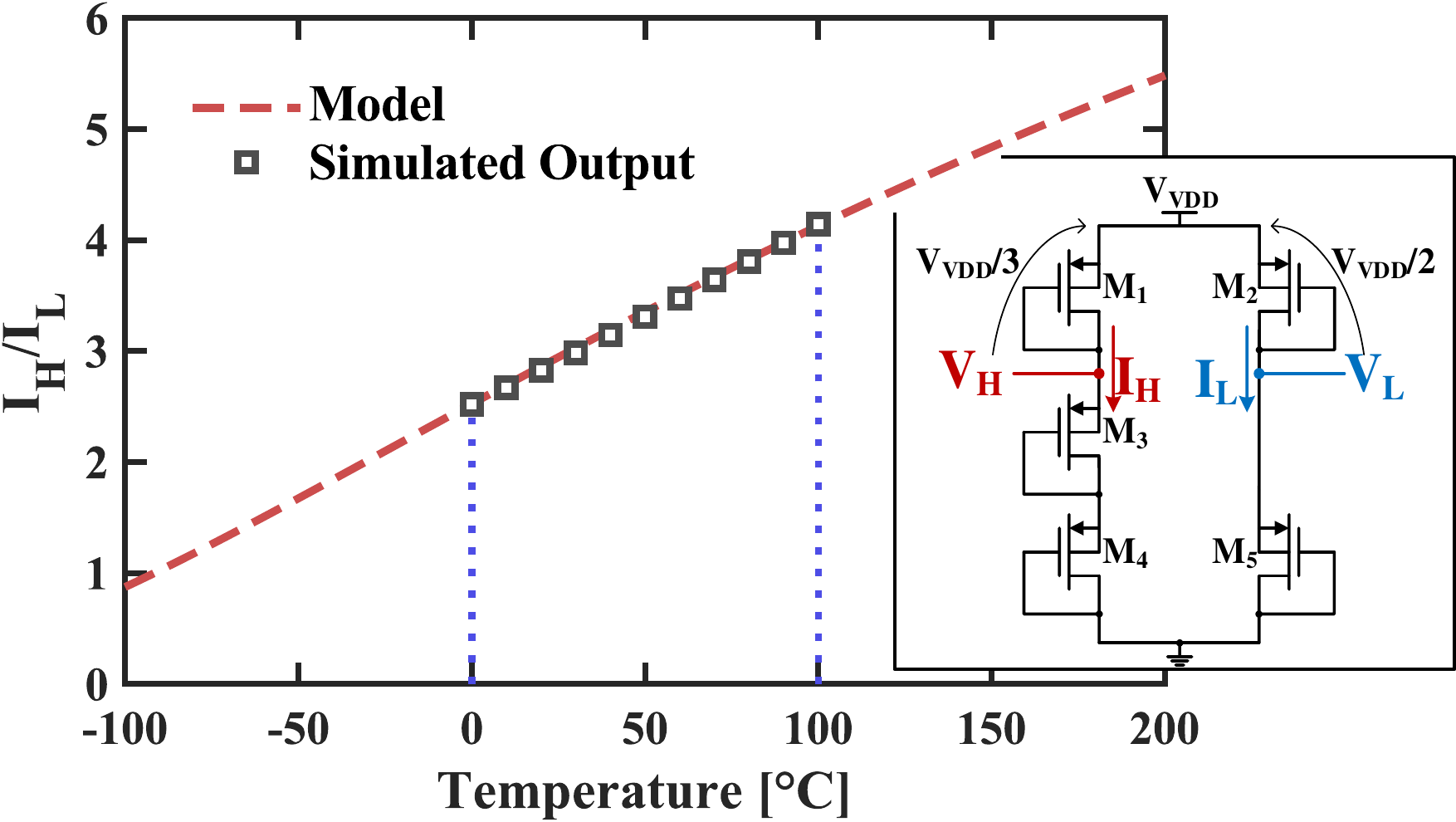}\vspace{-3mm}
     \caption{Simulated  $I_{H}/I_{L}$ current ratio as a function of the temperature.  In the inset: the temperature-to-current converter (TCC) circuit.}
     \label{fig:FIG2}
     \vspace{-5mm}
   \end{figure}

The proportional-to-absolute-temperature (PTAT) behavior of the current ratio $I_{H}/I_{L}$ is obtained thanks to the low-complexity circuit shown in the inset of \figref{fig:FIG2}. Such circuit is organized in two branches, including only five diode-connected PMOS devices. The left (right) branch is implemented with 3 (2) identically sized transistors, so that the $V_{VDD}$ is equally partitioned between them resulting in $V_{SG}=V_{VDD}/3$  ($V_{SG}=V_{VDD}/2$).
Since all the devices of the TCC operate in the sub-threshold region, with a source-to-drain voltage, $V_{SD}$, larger than four thermal voltages $V_T$ = $k_{B}T/q$ ($k_B$ is the Boltzmann constant, $T$ is the absolute temperature, and $q$ is the electron charge), their source current $I_{S}$ can be expressed as \cite{alioto2012ultra}: 
\begin{equation}
\vspace{0mm}
\label{eqn:eq1}
I_S = \frac{W}{L} I_0\,exp{\left( \frac{q(|V_{GS}| - |V_{th}|}{nk_{B}T}\right)},
\vspace{0mm}
\end{equation}
where $W/L$ is the aspect ratio of the device, $I_0$ is the technology dependent subthreshold current which can be obtained by extrapolating the current for $V_{GS} = V_{th}$, $V_{th}$ is the threshold voltage and $n$ is the subthreshold factor.%\\ $k_B$ is the Boltzmann constant and $T$ is the absolute temperature.%\\

Thus, the current ratio $I_{H}/I_{L}$ can be written as:
\begin{equation}
\vspace{0mm}
\label{eqn:eq4}
\frac{I_H }{I_L } =  \frac{ \left[ \frac{W}{L} \right]_1 }{ \left[ \frac{W}{L} \right]_2 }  \, exp{ \left(- \frac{q V_{VDD}}{ 6nk_{B} } \cdot \frac{1}{T}        \right)  },
\vspace{0mm}
\end{equation}
where we have assumed that $I_{0}$ and $V_{th}$ for transistors M1 and M2 are the same.

Based on the value assumed by the term $-qV_{VDD}/6nk_{B}$, the exponential can be well approximated by a linear relation in a bounded temperature range (e.g. 0\C -- 100\C), according to:
\begin{equation}
\vspace{0mm}
\label{eqn:eqX}
\frac{I_H }{I_L } \approx  m \times Temp + p,
\vspace{0mm}
\end{equation}
with $m$ = 16.1 $1/\C$, $p$ = 2.55 for our design, and $Temp$ expressed in \C. The obtained model equation \eqref{eqn:eq4} is plotted in \figref{fig:FIG2}, where it can be appreciated the good agreement with simulation results.

 \begin{figure}[t!] % [b]-> bottom, [t]->top, [H]->Here! ([h!] should do a better job),     {figure*}->float
     \centering
     %\vspace{-3mm}
     \includegraphics[width=0.40\textwidth]{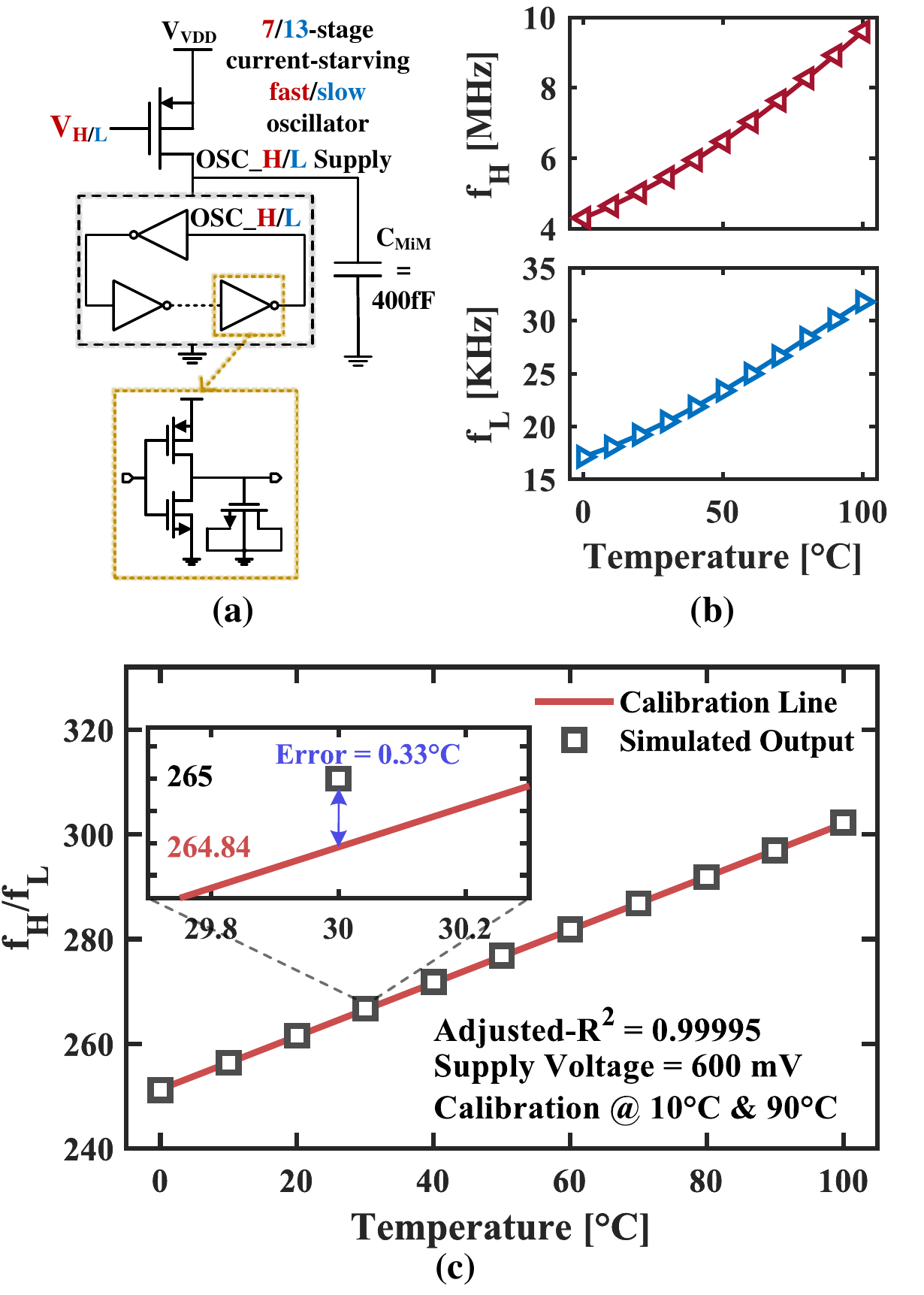}\vspace{-3mm}
     \caption{(a) Current-to-frequency converter (CFC), (b) simulated $f_H$ and $f_L$ vs. temperature  and (c)  simulated $f_H$/$f_L$ frequency ratio as a function of temperature in the \mbox{0\C--100\C} range, $V_{DD}$ = 600\mV.}
     \label{fig:FIG3}
     \vspace{-5mm}
   \end{figure}
   
\subsection{Current-to-Frequency Converter}\label{sec:CFC}  
As shown in \figref{fig:FIG3}(a), the CFC exploits two current-starved ring oscillators to convert the mirrored version of the  $I_{H}$ and $I_{L}$ currents into two digital pulse signals, whose frequency ratio $f_H$/$f_L$ maintains a linear PTAT trend. 

In general, the oscillation period $T_{osc}$ of a current-biased ring oscillator can be expressed as~\cite{someya201911}: 
\begin{equation}
\vspace{0mm}
\label{eqn:eq5}
T_{osc} = N \left(\frac{C_L \Delta V}{I_{bias}} + t_{fall} + t_{rise}\right) 
\vspace{0mm}
\end{equation}
where $N$ is the number of stages in the ring oscillator, $C_L$ is the load capacitance of a single delay cell, $\Delta V$ is the output voltage amplitude, $I_{bias}$ is the biasing current, while $t_{fall}$ and $t_{rise}$ are the falling and the rising times of a single stage. Both $t_{fall}$ and $t_{rise}$ should be sufficiently small compared with $(C_L \Delta V)/I_{bias}$ so that the oscillation frequency can be approximated as proportional to $I_{bias}/(C_L \Delta V)$. In this case,
$f_H$/$f_L$ ($=$ $T_{osc,L}/T_{osc,H}$) provides a similar PTAT characteristic
as $I_{H}$/$I_{L}$.

In our design, both the ring oscillators use the same delay cell topology (see \figref{fig:FIG3}(a)), based on a standard CMOS inverter loaded by a NMOS capacitor. The latter limits the oscillation frequency, while avoiding excessive increasing in the number of delay cells. As an additional benefit, the added capacitance at the output node of each inverter helps in increasing the linearity of the oscillation frequency with the biasing current (as from (4)). The  two  ring  oscillators have been designed with a proper number of stages (13 and 7 for the slow and the fast oscillators, respectively) and delay stage sizing. As shown in \figref{fig:FIG3}(a), decoupling MIM capacitors (400\fF) were introduced at the supply voltage nodes of the two oscillators to reduce the impact of the switching noise on the oscillation frequency.

Simulation results of the FDC circuit driven by the TCC are shown in \mbox{\figref{fig:FIG3}(b-c). From \figref{fig:FIG3}(b)}, as temperature increases from \mbox{0\C to 100\C, $f_L$ ($f_H$) spans from 17\kHz(4.3\MHz)} to \mbox{31.8\kHz(9.6\MHz)}. As shown in \mbox{\figref{fig:FIG3}(c)}, the adjusted-$R^2$ of 0.99995, evaluated on the $f_H$/$f_L$ ratio, leads to a tolerable error after a two-point calibration, with \mbox{10\C and 90\C} as temperature references.

\subsection{Voltage Regulation}\label{sec:VddStack}
\begin{figure}[!b] % [b]-> bottom, [t]->top, [H]->Here! ([h!] should do a better job),     {figure*}->float
  \vspace{-3mm}
     \centering
     \includegraphics[width=0.40\textwidth]{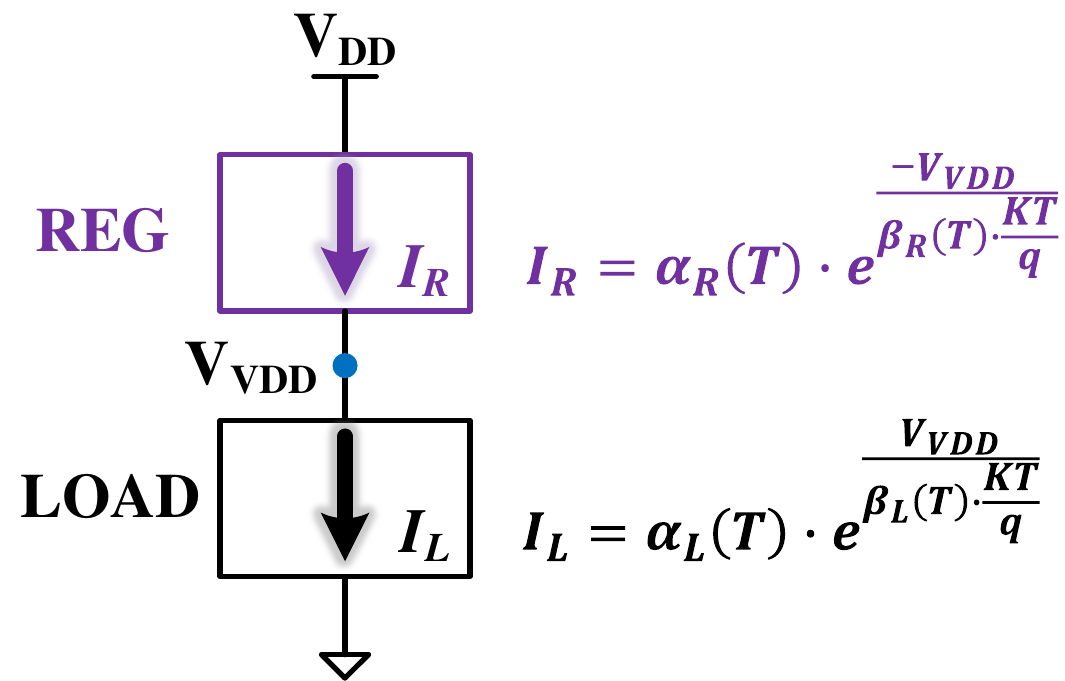}\vspace{-3mm}
     \caption{Regulator and loading circuits model equations for the temperature sensitivity characterization of the $V_{VDD}$.}
     \label{fig:FIG4}
     %\vspace{-5mm}
   \end{figure}

A low-complexity voltage regulation mechanism has been implemented to improve line and temperature sensitivities of TCC and CFC blocks. This has been achieved by interposing two series-connected native NMOS devices between $V_{DD}$ and the TCC+CFC circuits, as depicted in \figref{fig:FIG1}. In this way, the source of the lower native transistor corresponds to the virtual supply voltage $V_{VDD}$ seen by the TCC and CFC blocks. Note that, the stack of two native transistors will always be operated in sub-threshold, given that both have $V_{G}=V_{th}=0V$ and $V_{S}>0V$. The use of two long channel transistors in series to implement the voltage regulator block, makes its $I(V,T)$ characteristic essentially insensitive to variations of the external $V_{DD}$. Moreover, such approach also leads to an almost stable $V_{VDD}$ with temperature variations. In order to better explain this, we rely on the scheme reported in \figref{fig:FIG4}, where the regulator (REG) and the TCC and CFC circuits (LOAD) are modeled independently. For the regulator, an $I(V,T)$ characteristic in the following form is assumed:
\begin{equation}
\vspace{0mm}
\label{eqn:eq6}
I_{R} = \alpha_{R}(T)\,exp{\left( \frac{-q\, V_{VDD}}{\beta_{R}(T)\,k_{B}T}\right)},
\vspace{0mm}
\end{equation}
in which the DIBL effect is neglected (i.e., the $V_{DS}$ impact on $I_{D}$).

Such a kind of regulator is capable to provide an almost constant $V_{VDD}$ to a load a circuit whose $I(V,T)$ relation can be expressed in the following form: 
\begin{equation}
\vspace{0mm}
\label{eqn:eq7}
I_{L} = \alpha_{L}(T)\,exp{\left( \frac{q\, V_{VDD}}{\beta_{L}(T)\, k_{B}T}\right)},
\vspace{0mm}
\end{equation}
which is reasonable for CMOS circuits operated in sub-threshold regime, such as the TCC and CFC blocks in our design.

 \begin{figure}[!t] % [b]-> bottom, [t]->top, [H]->Here! ([h!] should do a better job),     {figure*}->float
     \centering
     \includegraphics[width=0.43\textwidth]{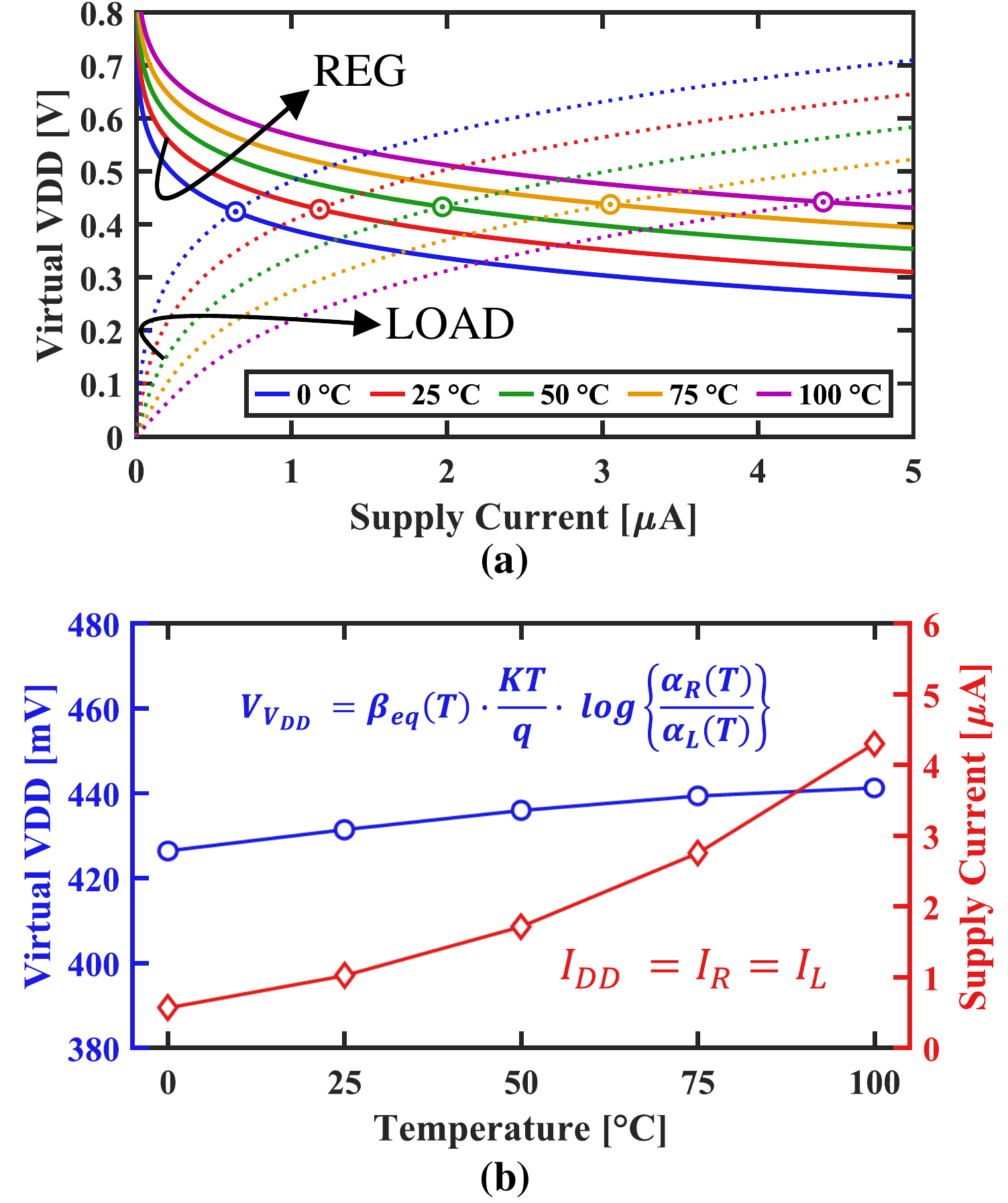}
     \vspace{-2mm}
     \caption{(a) Load curves of regulator and loading circuits at different temperatures. (b) Model equations of the $V_{VDD}$ and supply current drawn by the TCC and CFC blocks as a function of the temperature.}
     \label{fig:FIG5}
     \vspace{-5mm}
   \end{figure}

In \figref{fig:FIG5}(a), the DC V-I curves for regulator and load,  simulated at different temperatures, are depicted. The analytical expression of these curves can be easily obtained by writing \eqref{eqn:eq6} and \eqref{eqn:eq7} in their respective logarithmic forms.
The crossing points between the two curve families represent the load conditions for different temperatures. From this plot, it can be easily understood how the $V_{VDD}$ is essentially stable as temperature changes, despite the supply current exponentially increases with temperature.

\begin{figure}[!t] % [b]-> bottom, [t]->top, [H]->Here! ([h!] should do a better job),     {figure*}->float
     \centering
    % \vspace{-5mm}
     \includegraphics[width=0.46\textwidth]{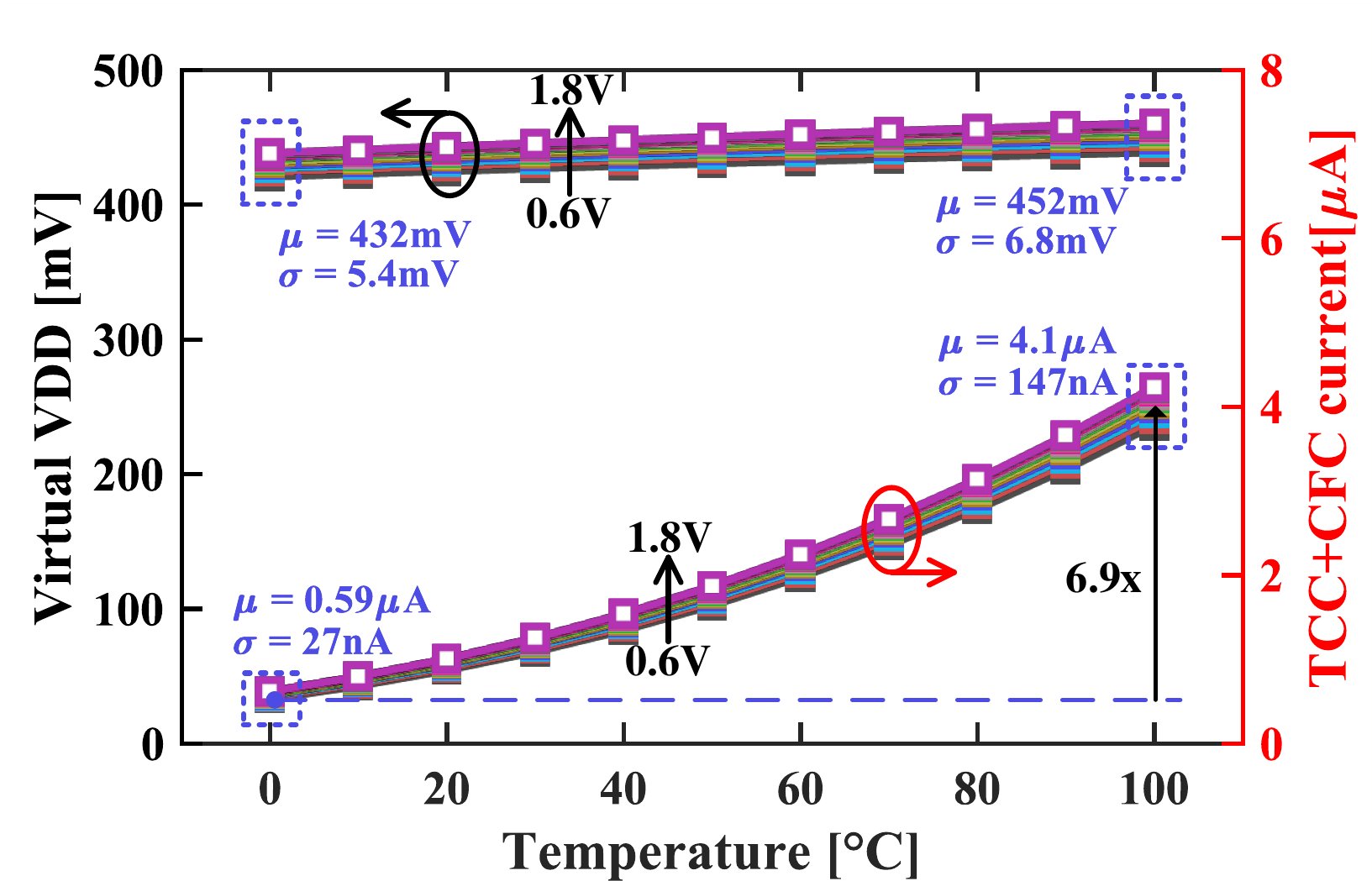}\vspace{-3mm}     \caption{Simulated Virtual $V_{DD}$ and temperature-to-current converter $+$ current-to-frequency converter drawn current for supply voltage ranging from 0.6\V to 1.8\V.}
     \label{fig:FIG6}
     \vspace{-5mm}
   \end{figure}

Analytically, by manipulating \eqref{eqn:eq6} and \eqref{eqn:eq7}, one can put the $V_{VDD}$  in the following form:
\begin{equation}
\vspace{0mm}
\label{eqn:eq8}
V_{VDD} =  \beta_{eq}(T)\,\frac{k_{B}T}{q}\,log{\left(\frac{\alpha_{R}(T)}{\alpha_{L}(T)}\right)}
\vspace{0mm}
\end{equation}

\begin{equation}
\vspace{0mm}
\label{eqn:eq9}
\text{with}\,\,\beta_{eq} =  \frac{\beta_{L}\,\beta_{R}}{\beta_{L}+\beta_{R}}\,,
\vspace{0mm}
\end{equation}
in which all the terms  $\beta_{eq}$, $\frac{k_{B}T}{q}$,  $\alpha_{R}$ and $\alpha_{L}$ are individually a function of the temperature. In fact, we have verified that the $\frac{\alpha_{R}}{\alpha_{L}}$ ratio is essentially constant, while the $\beta_{eq}$ term has an almost 1/T dependence which compensates the linear term in $\frac{k_{B}T}{q}$. The aforementioned fitting parameters have been extracted for the regulator and load circuits at the different temperatures, and the corresponding analytical expression for $V_{VDD}$ and the supply current are reported in \figref{fig:FIG5}(b): when the temperature raises from 0\C to 100\C, although an about 7 times current increase is observed, the corresponding change in the voltage is of about 3\%. The extracted model equations are consistent with the sensor front-end simulations reported in \figref{fig:FIG6}, which shows the behavior of  $V_{VDD}$ voltage node and current drawn by the front-end (stacked native NMOS devices biasing TCC and CFC) sensor circuitry as a function of temperature for $V_{DD}$ ranging from 0.6\V to 1.8\V with a step of 100\mV. It can be easily observed that while the average current increases from about 0.6\uA @ T= 0\C to 4.1\uA @ T= 100\C (i.e., drawn current increases more than 6.9 times as temperature passes from 0\C to 100\C), the $V_{VDD}$ increases of less than 5\% in the same temperature range. At the same time, $V_{VDD}$ stays quite constant for $V_{DD}$ spreading from 0.6\V to 1.8\V, as a consequence of long channel stacked NMOS devices used to screen the $V_{DS}$ dependence. From the above considerations, are then evident the benefits of the adopted low-complexity voltage regulation to improve both line and temperature sensitivities.

In deeply scaled process nodes (i.e. below 40-nm), native MOSFETs could not be available. In this case, the described voltage regulation can be realized by properly sized nMOS devices with regular $V_{th}$ (RVT), as long as they are biased with a stable gate voltage close to the $V_{th}$, so that the ${V_S > 0}$ leads them to operate in the sub-threshold regime.
 
 \begin{figure}[!b] % [b]-> bottom, [t]->top, [H]->Here! ([h!] should do a better job),     {figure*}->float
     \centering \vspace{-3mm}
     \includegraphics[width=0.49\textwidth]{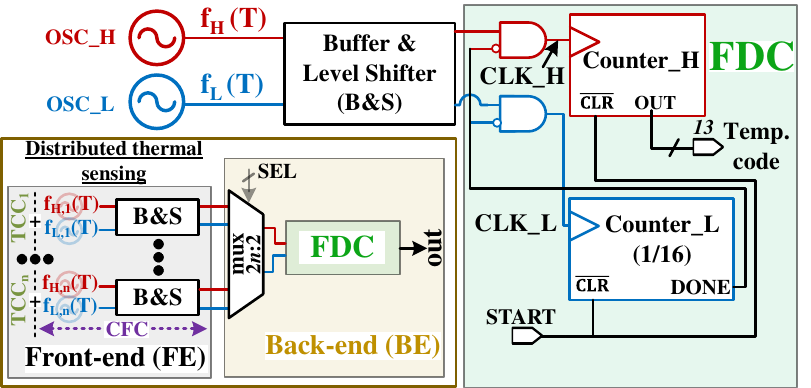} %FIG7
     \caption{Frequency-to-digital converter architecture. The inset shows the principle scheme for distributed thermal sensing.}
     \label{fig:FIG7}
     \vspace{-0mm}
   \end{figure}

\subsection{Frequency-to-Digital Converter}\label{sec:FDC}
The principle scheme of the digital back-end, responsible for generating the digital PTAT code, is shown in \figref{fig:FIG7}. Since the amplitude of the signals produced by the two ring oscillators in the CFC is constrained by the $V_{VDD}$ voltage level, both $OSC_{H,L}$ signals are up-converted to the $V_{DD}$ voltage domain by the compact and energy-efficient level shifter (LS) proposed in~\cite{lanuzza2016ultralow}. Such circuit exhibits a very large voltage conversion range, and it is adaptive to change in $V_{DD}$ voltage level~\cite{lanuzza2016ultralow}, rendering the whole temperature sensor fully compatible with  DVFS systems. Frequency to digital conversion is then performed by two asynchronous counters.
The $Counter\_L$ acts as reference counter to set the time window of the temperature measurement sampling. This 5-bit counter defines a sampling time window of 16/$F_L$ seconds, since only 4 bits are used for counting while the MSB is exploited for triggering purpose. On the other side the resolution of the temperature sensor depends on $Counter\_H$, which was sized for 13-bit to prevent counting  overflow over the temperature detection range from \mbox{0\C to 100\C}, also considering some margin against undesired offsets due to process/mismatch variations. 

\begin{figure}[!t] % [b]-> bottom, [t]->top, [H]->Here! ([h!] should do a better job),     {figure*}->float
     \centering
     %\vspace{-3mm}
     \includegraphics[width=0.47\textwidth]{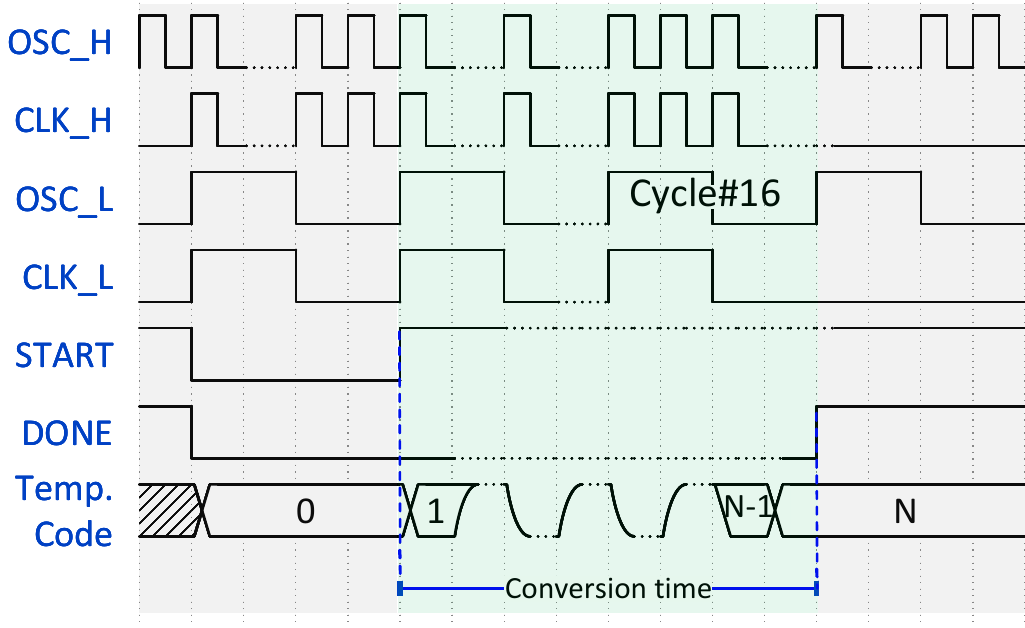}\vspace{-2mm}
     \caption{Timing diagram of the FDC block.}
     \label{fig:FIG8}
     \vspace{-4mm}
  \end{figure}

The timing diagram of the FDC block is shown in \mbox{\figref{fig:FIG8}}. When the START signal is  triggered (note that, the START and the $OSC_{L}$ signals are synchronized), both counters begin counting upward until the transition to '1' of the DONE signal (i.e., the MSB of the reference counter), which happens after  sixteen cycles of the  slower oscillator. At this time, a temperature code is available at the output of the $Counter\_H$. Both the counters are reset by the START signal when a new temperature measurement is required.

 \begin{figure}[!t] % [b]-> bottom, [t]->top, [H]->Here! ([h!] should do a better job),     {figure*}->float
     \centering
     \includegraphics[width=0.46\textwidth]{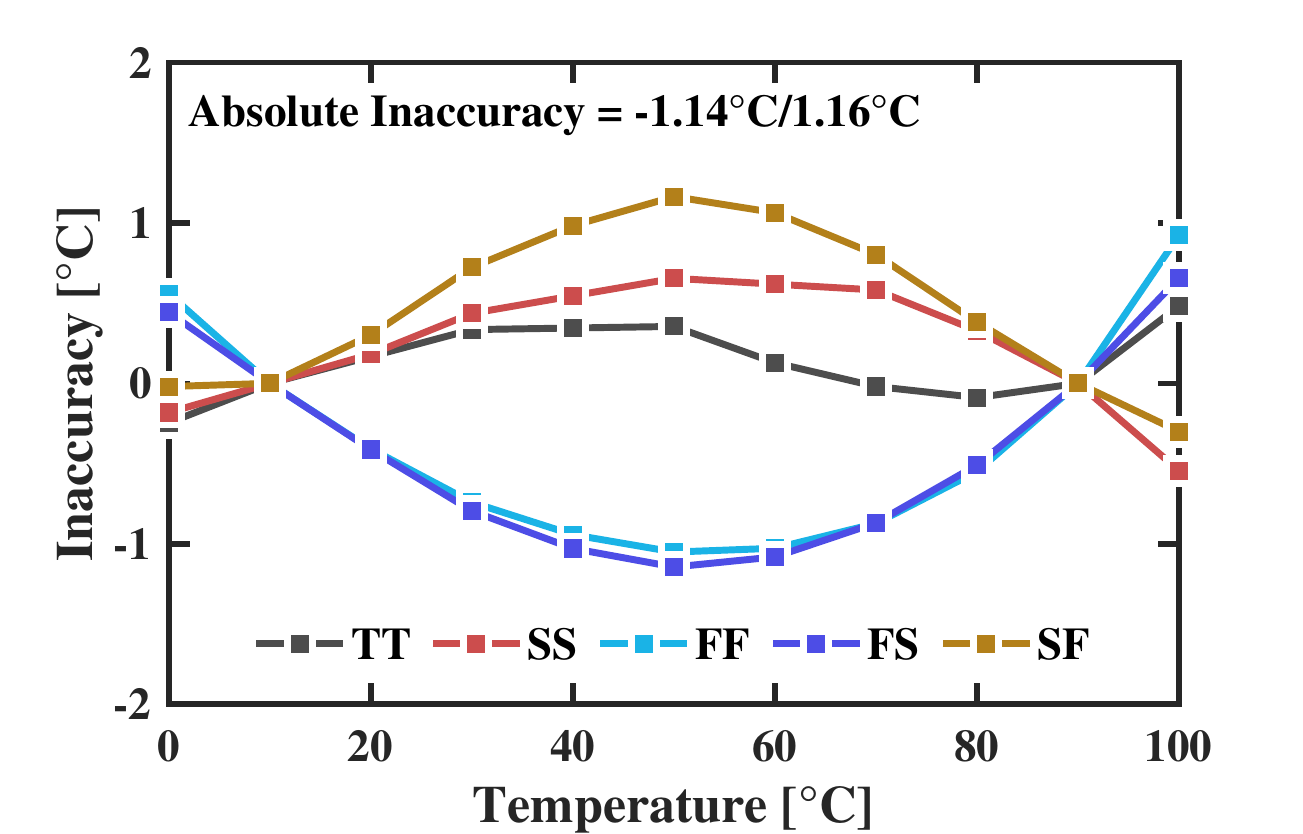}
     \caption{Simulated inaccuracy as a function of temperature for different process corners. Temperature calibration points: 10\C and 90\C.}
     \label{fig:FIG9}
     \vspace{-5mm}
   \end{figure}

 \begin{figure}[!b] % [b]-> bottom, [t]->top, [H]->Here! ([h!] should do a better job),     {figure*}->float
     \centering \vspace{-4mm}
     \includegraphics[width=0.46\textwidth]{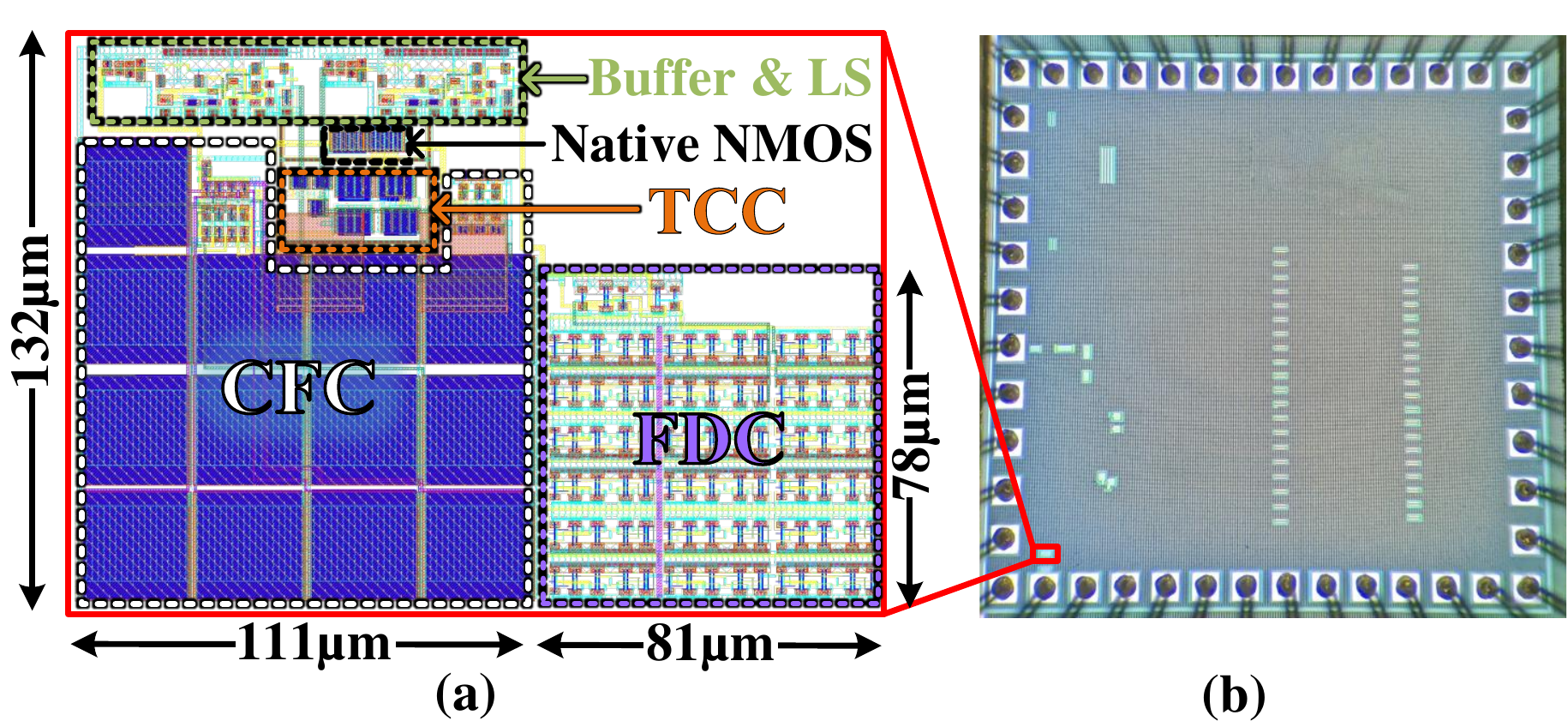}
     \caption{(a) Layout of the designed temperature sensor and (b) micrograph of the test chip.}
     \label{fig:FIG10}
     \vspace{-0mm}
   \end{figure}

In a general system, the FDC circuit can be shared between multiple sensor front-ends as shown in the sketch in \figref{fig:FIG7}. In this case, the back-end comprises two n-to-1 multiplexers, which take n output pairs from different sensor front-ends (i.e. each pair consists of the two signals at $f_H$ and $f_L$ frequencies) and pass the selected pair to the shared FDC. This allows to save precious silicon area when multiple thermal information is required for an effective DTM.

   \begin{figure*}[t] % [b]-> bottom, [t]->top, [H]->Here! ([h!] should do a better job),     {figure*}->float
     \centering
     \includegraphics[width=0.90\textwidth]{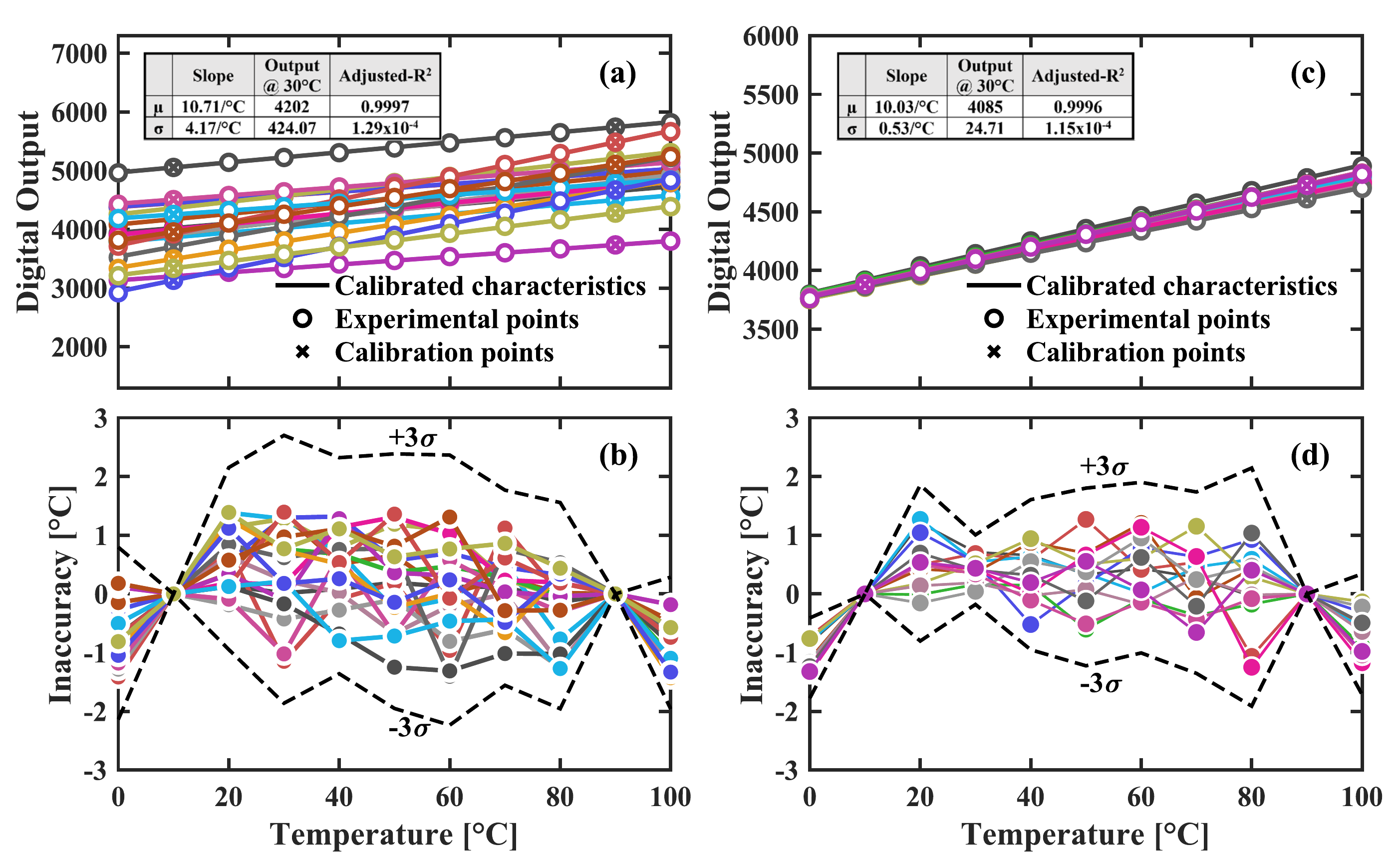} \vspace{-3mm}
     \caption{Measured (a) digital output and (b) corresponding inaccuracy as a function of the temperature for $V_{DD}$ = 0.6\V in 20 test chips. Measured (c) digital output and (d) inaccuracy for a typical sample as a function of the temperature over the 0.6\V -- 1.8\V supply voltage range.}
     \label{fig:FIG11}
     \vspace{-4mm}
   \end{figure*} 

\figref{fig:FIG9} reports simulated inaccuracy for the output temperature code in the target temperature detection range while considering different process corners and 
two-point calibration with 10 \C and 90 \C as temperature references. An absolute inaccuracy of --1.14 \C $/$ 1.16 \C is observed at 50 \C for the FS/SF corners. Such result suggests a low susceptibility of the proposed design against process variability.

\vspace{-2mm}
\section{Experimental Results}\label{sec:Results}
The fully-integrated temperature sensor was fabricated in a 180\nm CMOS technology node, while occupying a very small silicon area (less than 0.021\mmsquared). The physical design is shown in  \figref{fig:FIG10}(a), whereas a micrograph of the test chip is reported in \figref{fig:FIG10}(b). A silicon area of 14650\umsquared is occupied by the voltage-regulated sensor front-end, while the FDC footprint requires about 6300\umsquared.

Twenty test chips were measured in the target temperature range considering two-point calibration, with 10\C and 90\C as temperature references. Obtained results are summarized in \figref{fig:FIG11}. More precisely, \figref{fig:FIG11}(a) reports the temperature code of measured chips calibrated for $V_{DD}$=600\mV. A very good linearity was found for all the samples with an average adjusted-$R^2$ of 0.9997, which is only slightly degraded with respect to that observed from the simulation results (see \figref{fig:FIG3}(c)). However, from \mbox{\figref{fig:FIG11}}(a), a noticeable difference can be appreciated in the slope coefficient from sample to sample. This prevents a simple 1-point calibration to be exploited when significantly high accuracy in the measurement must be guaranteed. The inaccuracy of the measurements as a function of the temperature is shown in \figref{fig:FIG11}(b). For all the measured samples, the inaccuracy is always maintained within the --1.45\C/1.4\C range. Also looking to the pessimistic {3$\sigma$} data (about $\pm$ 2.5\C), such measurement results are well within the acceptable accuracy specs for an effective DTM in state-of-the-art SoC~\cite{vinshtok2020ultra,li2009}. Additional measurements performed on a single sensor are reported in \figref{fig:FIG11}(c-d). Here, the effect of voltage scaling was evaluated by considering a sweep on the $V_{VDD}$ from 0.6\V to 1.8\V with 100\mV step. Note that, for each supply voltage, an independent two-point calibration was considered. From \figref{fig:FIG11}(c), it can be observed that the digital temperature code differs significantly only at the highest temperatures, whereas, as shown in \figref{fig:FIG11}(d), the inaccuracy spans from --1.35\C @ $V_{DD}$ = 1.5\V to 1.27\C @ $V_{DD}$ = 0.8\V. The {3$\sigma$} inaccuracy is within the $\pm$ 2\C range.

\begin{figure}[!t] % [b]-> bottom, [t]->top, [H]->Here! ([h!] should do a better job),     {figure*}->float
     \centering
     \includegraphics[width=0.45\textwidth]{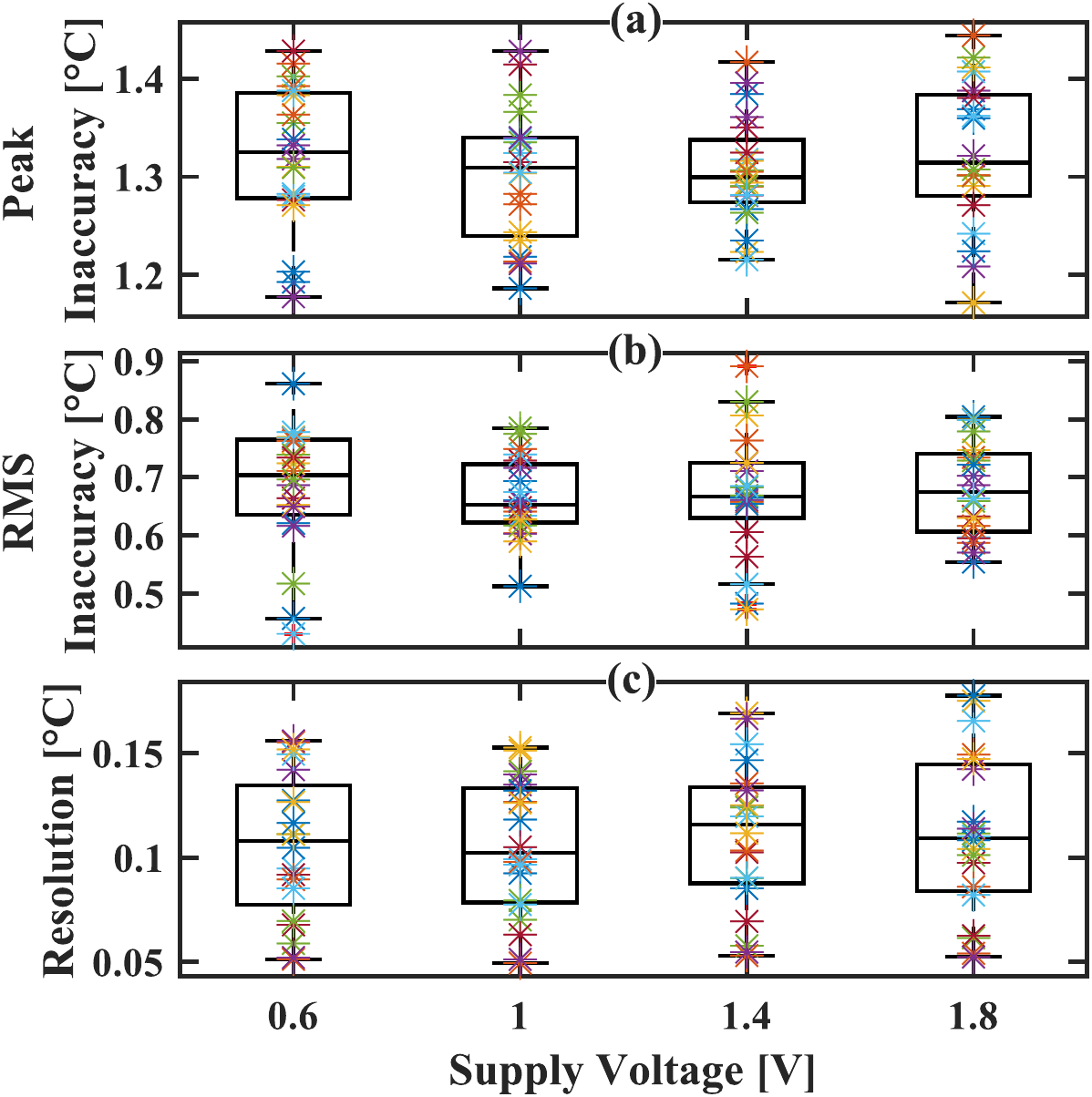}
          \vspace{2mm}
     \caption{Measured peak inaccuracy (a), RMS inaccuracy (b) and resolution (C) for different supply voltages and 20 test chips.}
     \label{fig:FIG12}
     \vspace{-2mm}
   \end{figure}

\begin{figure}[t] % [b]-> bottom, [t]->top, [H]->Here! ([h!] should do a better job),     {figure*}->float
     \centering
     %\vspace{-3mm}
     \includegraphics[width=0.45\textwidth]{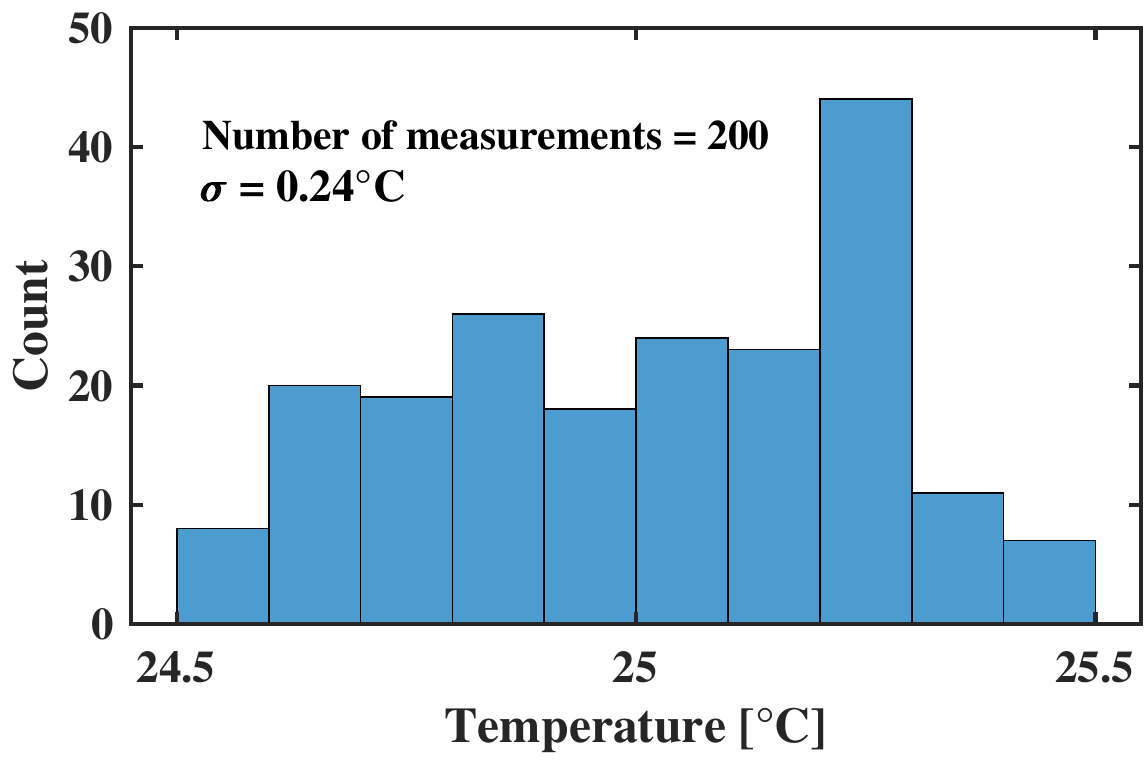}\vspace{-2mm}
     \caption{Measured histogram of 200 different temperature readings at temperature $T$ =\mbox{25\C.}}
     \label{fig:FIG13}
     \vspace{-6mm}
  \end{figure}

Die-to-die variability in the target temperature detection range is fully evaluated in \figref{fig:FIG12}(a-c), by considering all test samples and 4 different supply voltage values (i.e., $V_{DD}$=0.6\V, 1\V, 1.4\V and 1.8\V). From \figref{fig:FIG12}(a), the peak inaccuracy is within a range of 1.1\C $-$ 1.45\C, with a median value close to 1.3\C. This corresponds to a RMS inaccuracy in the range from 0.42\C to 0.89\C, as shown in \figref{fig:FIG12}(b). Whereas the achieved resolution (i.e., the inverse slope of the sensor output characteristics) shows very competitive values with a worst case of 0.22\C for $V_{DD}$ = 1.8\V, as reported in \figref{fig:FIG12}(c). The measured average resolution is 0.132\C, 0.131\C, 0.138\C and 0.139\C for $V_{DD}$ equal to 0.6\V, 1\V, 1.4\V and 1.8\V, respectively. Indeed, the sampled temperature code can change in each measurement due to thermal noise. To properly evaluate the sensor resolution, the temperature code was measured 200 times at the fixed
temperature of \mbox{25\C} for a typical sample. \mbox{\figref{fig:FIG13}} reports the result of this experiment. The standard deviation {$\sigma$} of the samples is \mbox{0.24\C} (i.e. the noise-limited resolution), which corresponds to 1.84 LSB.

\begin{figure}[h!] % [b]-> bottom, [t]->top, [H]->Here! ([h!] should do a better job),     {figure*}->float
     \centering
     \includegraphics[width=0.45\textwidth]{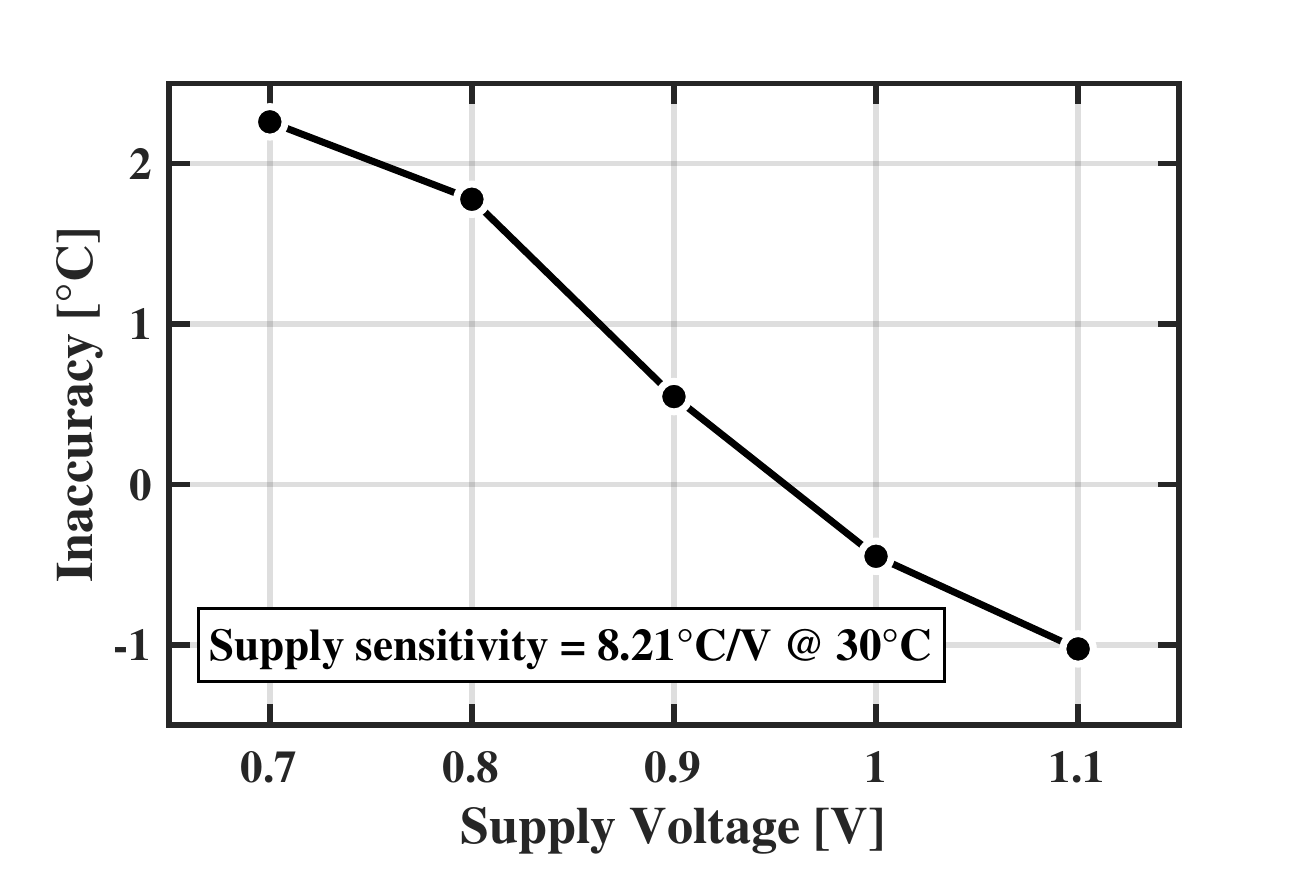}
     \vspace{-3mm}
     \caption{Measured inaccuracy vs supply voltage at temperature $T$ = 30\C. For this measurement the sensor was calibrated only at 0.9\V.}
     \label{fig:FIG14}
     \vspace{-2mm}
   \end{figure}

Sensitivity to unwanted voltage variations for the typical sample is characterized in \figref{fig:FIG14}, which shows inaccuracy as a function of supply voltage. A line sensitivity of 8.21\C/\V at a temperature of 30\C was measured by calibrating the sensor only at 0.9\V and then extracting the inaccuracy when supply changes in the $\pm$200\mV (i.e., $\pm$22\%) range.  However, when supply voltage of the sensor is changed as an effect of DVFS adjustment, the line sensitivity can be further reduced down to 0.56\C/\V at 30\C. In fact, in this case, the two temperature calibration points for each possible $V_{DD}$ (i.e. within the chip operating $V_{DD}$ range) can be easily pre-stored to define the actual calibrated characteristic (see \figref{fig:FIG11}(d)). 

\begin{figure}[h!] % [b]-> bottom, [t]->top, [H]->Here! ([h!] should do a better job),     {figure*}->float
     \centering
     \includegraphics[width=0.45\textwidth]{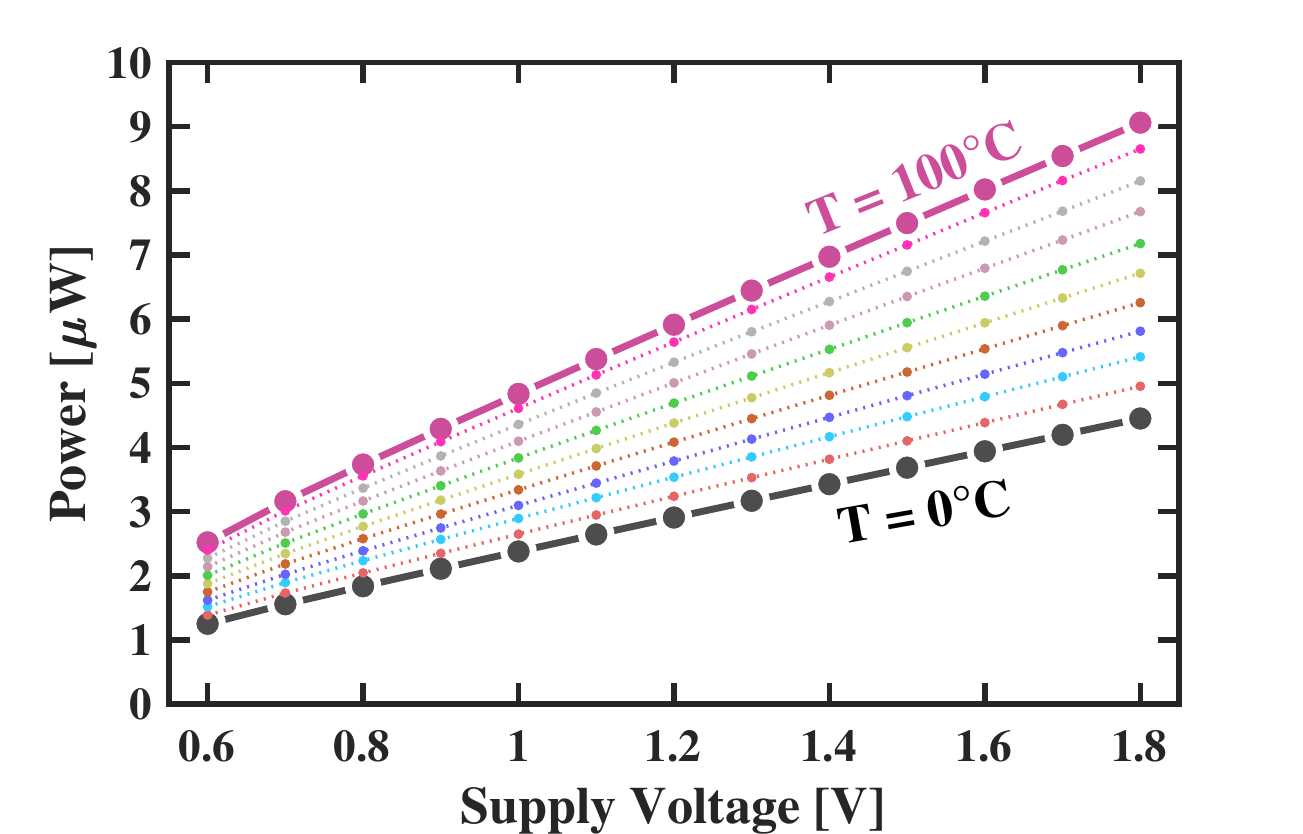}
     \vspace{-3mm}
     \caption{Measured power as  a function of supply voltage for temperature spreading from 0\C to 100\C.}
     \label{fig:FIG15}
     \vspace{-6mm}
   \end{figure}

Finally, \figref{fig:FIG15} shows the power consumption of the proposed sensor as a function of $V_{DD}$ for temperature ranging from 0\C to 100\C. Power increases almost linearly with the $V_{DD}$, spreading from 1.57\uW ($V_{DD}$ = 0.6\V) to 5.61\uW ($V_{DD}$ = 1.8\V) at 25\C. On the other side, power also increases with the temperature, reaching a maximum value of about 9\uW ($V_{DD}$= 1.8\V and $Temp$= 100\C).

\section{Comparison} \label{sec:comparison}
Table \ref{table} summarizes measurement results of the temperature sensor as compared to alternative CMOS designs~\cite{anand2016vco_JSSC16, wang2019763_JSSC19_763p, someya201911,li2020a+_TCASI21, vinshtok2020ultra, zambrano20210, huang2017energy, an2014energy, Tang2018, someya2020_6.4n, tang2020leak}. Experimental results provided for our design come from measurements on 20 test chips, while that of most of the competitors are based on a smaller number of samples (except for~\cite{vinshtok2020ultra} and~\cite{ tang2020leak}). The circuits reported in~\cite{ vinshtok2020ultra, huang2017energy, an2014energy, Tang2018} do not support operation with sub 1\V voltages, while sensors discussed in~\cite{anand2016vco_JSSC16, zambrano20210, huang2017energy, an2014energy} show poor voltage scalability. On the contrary, our sensor supports the larger supply voltage operating range from 0.6\V to 1.8\V, while exhibiting a supply sensitivity of about 8 \mbox{\C/\V}. In addition, it does not require any external reference signal unlike the circuits reported in~\mbox{\cite{someya2020_6.4n} and ~\cite{tang2020leak}}, which use a reference clock~\mbox{\cite{tang2020leak}} and a reference voltage~\mbox{\cite{someya2020_6.4n}}, respectively. Moreover, the proposed design exhibits a relatively small footprint of just 0.021\mmsquared. Indeed, it turns out to be the less area-hungry design in 180\nm and the third in absolute terms.
\begin{table*}[t]
\centering
\setlength{\tabcolsep}{0.15em} %padding space
\caption{COMPARISON WITH THE STATE-OF-THE-ART}
\label{table}

\begin{threeparttable}  %enable table footnotes
\scriptsize             %letter size
\begin{tabular}{@{}ccccccccccccc@{}}
\toprule
\multicolumn{1}{l}{\textbf{}} &
  \textbf{\begin{tabular}[c]{@{}c@{}}SENSORS'14\\ \cite{an2014energy}\end{tabular}} &
\textbf{\begin{tabular}[c]{@{}c@{}}JSSC'16\\ \cite{anand2016vco_JSSC16}\end{tabular}} &
  \textbf{\begin{tabular}[c]{@{}c@{}}SENSORS'17\\ \cite{huang2017energy}\end{tabular}} &
  \textbf{\begin{tabular}[c]{@{}c@{}}SENSORS'18\\ \cite{Tang2018}\end{tabular}} &
  \textbf{\begin{tabular}[c]{@{}c@{}}JSSC'19\\ \cite{wang2019763_JSSC19_763p}\end{tabular}} &
  \textbf{\begin{tabular}[c]{@{}c@{}}JSSC'19\\ \cite{someya201911}\end{tabular}} &
  \textbf{\begin{tabular}[c]{@{}c@{}}TCASI'21\\ \cite{li2020a+_TCASI21}\end{tabular}} &
  \textbf{\begin{tabular}[c]{@{}c@{}}ACCESS’20\\ \cite{vinshtok2020ultra}\end{tabular}} &
  \textbf{\begin{tabular}[c]{@{}c@{}}JSSC'20$^{\textbf{*}}$\\ \cite{tang2020leak}\end{tabular}} &
  \textbf{\begin{tabular}[c]{@{}c@{}}SSC-L'20$^{\textbf{*}}$\\ \cite{someya2020_6.4n}\end{tabular}} &
  \textbf{\begin{tabular}[c]{@{}c@{}}TCAS-II'21\\ \cite{zambrano20210}\end{tabular}} &
  \textbf{This work} \\ \midrule
\multicolumn{1}{c|}{\textbf{Technology {[}nm{]}}} &
  130 &
  65 &
  180 &
  130 &
  65 &
  180 &
  130 &
  65 &
  55 &
  65 &
  180 &
  180 \\
\multicolumn{1}{c|}{\textbf{Measured   Samples}} &
  7 &
  7 &
  8 &
  10 &
  12 &
  9 &
  9 &
  35 &
  64 &
  9 &
  9 &
  20 \\
\multicolumn{1}{c|}{\textbf{\begin{tabular}[c]{@{}c@{}}Voltage Scalability\\ (Voltage Operating\\ Range {[}V{]})\end{tabular}}} &
  No &
  \begin{tabular}[c]{@{}c@{}}Yes\\ (0.85 – 1.05)\end{tabular} &
  No &
  \begin{tabular}[c]{@{}c@{}}Yes\\ (1.05   - 1.4)\end{tabular} &
  \begin{tabular}[c]{@{}c@{}}Yes\\ (0.5 – 1)\end{tabular} &
  Yes &
  No &
  \begin{tabular}[c]{@{}c@{}}Yes\\ (1 – 1.5)\end{tabular} &
  \begin{tabular}[c]{@{}c@{}}Yes\\ (0.8 – 1.3)\end{tabular} &
  \begin{tabular}[c]{@{}c@{}}Yes\\ (0.7 – 1.05)\end{tabular} &
  No &
  \begin{tabular}[c]{@{}c@{}}Yes\\ (0.6 – 1.8)\end{tabular} \\
\multicolumn{1}{c|}{\textbf{\begin{tabular}[c]{@{}c@{}}Reference\\ Supply Voltage {[}V{]}\end{tabular}}} &
  1.2 &
  1 &
  1.8 &
  1.2 &
  0.5 &
  0.8 &
  0.95 &
  1.2 &
  0.9 &
  0.8 &
  0.35 &
  0.6 \\
  \multicolumn{1}{c|}{\textbf{Area {[}mm$^2${]}}} &
  0.031 &
  \begin{tabular}[c]{@{}c@{}}0.0082\\ (0.004 +\\ 0.0042$^{(a)}$)\end{tabular} &
  0.118 &
  0.06 &
  0.63 &
  0.074 &
  0.07 &
  0.0019 &
  \begin{tabular}[c]{@{}c@{}}0.0388\\ (0.0018 +\\ 0.037$^{(a)}$)\end{tabular} &
  0.32 &
  0.049 &
  \begin{tabular}[c]{@{}c@{}}0.021\\ (FE: 0.0147\\ BE:0.0063)\end{tabular} \\
\multicolumn{1}{c|}{\textbf{\begin{tabular}[c]{@{}c@{}}Temperature\\ range {[}°C{]}\end{tabular}}} &
  20 – 120 &
  0 – 100 &
  -20   – 120 &
  -20   – 100 &
  0 –   100 &
  -20 – 80 &
  0 –   80 &
  -10 – 100 &
  -40 – 125 &
  -30 – 70 &
  0 – 100 &
  0 –   100 \\
\multicolumn{1}{c|}{\textbf{\begin{tabular}[c]{@{}c@{}}Off-Chip\\ non-linearity\\ correction\end{tabular}}} &
  Yes &
  Yes &
  No &
  No (Yes) &
  No &
  No &
  Yes &
  Yes &
  Yes &
  No &
  No &
  No \\
\multicolumn{1}{c|}{\textbf{Calibration}} &
  1-point &
  2-point &
  2-point &
  1-point &
  2-point &
  2-point &
  2-point &
  2-point &
  2-point &
  2-point &
  2-point &
  2-point \\
\multicolumn{1}{c|}{{\color[HTML]{000000} \textbf{\begin{tabular}[c]{@{}c@{}}Min/Max\\ Inaccuracy   \mbox{{[}°C{]}}\end{tabular}}}} &
  -0.63/1.04 &
  -0.9/0.9 &
  -1.5/1.71 &
  \begin{tabular}[c]{@{}c@{}}-2.88/2.71\\ (-1.7/1.26)\end{tabular} &
  -1.53/1.61  &
  -0.9/1.2 &
  -0.4/0.44 &
  -1.62/2.04 &
  -0.37/0.72 &
  -1/0.7 &
  -3/3 &
  -1.45/1.4 \\
\multicolumn{1}{c|}{\textbf{\begin{tabular}[c]{@{}c@{}}Relative\\ Inaccuracy$^{\textbf{(b)}}$ {[}\%{]}\end{tabular}}} &
  1.67 &
  1.8 &
  2.86 &
  4.66 (2.55) &
  3.14 &
  2.1 &
  1.05 &
  4.54 &
  0.66 &
  1.7 &
  6 &
  2.85 \\
\multicolumn{1}{c|}{\textbf{Resolution {[}°C{]}}} &
  0.595$^{(e)}$ &
  0.3$^{(f)}$ &
  0.048$^{(f)}$ &
  0.34$^{(e)}$ &
  0.3$^{(e)}$ &
  0.145$^{(f)}$ &
  0.1$^{(e)}$ &
  0.32$^{(f)}$ &
  0.013$^{(f)}$ &
  0.075$^{(e)}$ &
  0.27$^{(e)}$ &
  \begin{tabular}[c]{@{}c@{}}0.13$^{(e)}$\\ 0.24$^{(f)}$\end{tabular} \\
\multicolumn{1}{c|}{\textbf{\begin{tabular}[c]{@{}c@{}}Conversion \\ Time {[}ms{]}\end{tabular}}} &
  0.0023 &
  0.022 &
  1 &
  0.0133 &
  300 &
  839 &
  59 &
  0.01 &
  1.3 &
  765 &
  33 &
  \begin{tabular}[c]{@{}c@{}}$\mu$ = 0.67\\ $\sigma$ = 0.081\end{tabular} \\
\multicolumn{1}{c|}{\textbf{\begin{tabular}[c]{@{}c@{}}Energy per\\ Conversion {[}nJ{]}\end{tabular}}} &
  0.67 &
  3.4 &
  93.6 &
  9.92 &
  0.23 &
  8.9 &
  11.56 &
  0.94 &
  12.8 &
  4.9 &
  0.46 &
  \begin{tabular}[c]{@{}c@{}}$\mu$ = 1.06\\ $\sigma$ = 0.127\end{tabular} \\
\multicolumn{1}{c|}{\textbf{R-FoM$^{\textbf{(c)}}$ {[}nJ$\cdot$K2{]}}} &
  0.237 &
  0.3 &
  0.216 &
  1.147 &
  0.02 &
  0.19 &
  0.116 &
  0.096 &
  0.022 &
  0.028 &
  0.033 &
  0.061\\
\multicolumn{1}{c|}{\textbf{\begin{tabular}[c]{@{}c@{}}Supply\\ sensitivity \mbox{[°C/V]}\end{tabular}}} &
  43$^{(d)}$ &
  34 &
  - &
  13.6 &
  8.4 &
  3.8 &
  13.7 &
  2.4 &
  5.76 &
  2.8 &
  16 &
  8.21 \\
\multicolumn{1}{c|}{\textbf{Power {[}W{]}}} &
  288µ &
  \begin{tabular}[c]{@{}c@{}}154µ\\ @27°C\end{tabular} &
  93.6µ &
  744µ &
  \begin{tabular}[c]{@{}c@{}}763p\\ @27°C\end{tabular} &
  \begin{tabular}[c]{@{}c@{}}11n\\ @25°C\end{tabular} &
  \begin{tabular}[c]{@{}c@{}}196n\\ @30°C\end{tabular} &
  94µ &
  9.3µ &
  6.4n &
  \begin{tabular}[c]{@{}c@{}}14n\\ @25°C\end{tabular} &
  \begin{tabular}[c]{@{}c@{}}1.57µ\\ @25°C\end{tabular} \\ \bottomrule
\end{tabular}
    
  \begin{tablenotes}
    \item[] (a) Estimated area for non-linearity correction logic \cite{anand2016vco_JSSC16,tang2020leak}.\,\,\,\,\,\,\,\,(c) R-FoM = Energy/Conversion × Resolution$^2$. \,\,\,\,\,\,\,(e) Counter resolution. \,\,\,\,\,* External reference signal required.
    \item[] (b) Relative Inaccuracy = (Max Inaccuracy – Min Inaccuracy)/Temperature Range × 100. \,\,\,\,\,\,\,\,\,\,\,\,\,\,\,\,\,\,\,\,\,\,\,\,\,\,\,\,\,\,\,\,\,\,\,\,\,\,\,\,\,\,\,\,\,\,\,\,\,\,\,\,\,\,\,\,\,\,\,\,(d) Simulated result.\,\,\,\,\,\,\,\,\,\,(f) Noise-limited resolution.
  \end{tablenotes}

\end{threeparttable}
\vspace{-6mm}
\end{table*}

\begin{figure}[!t] % [b]-> bottom, [t]->top, [H]->Here! ([h!] should do a better job),     {figure*}->float
     \centering
     \includegraphics[width=0.47\textwidth]{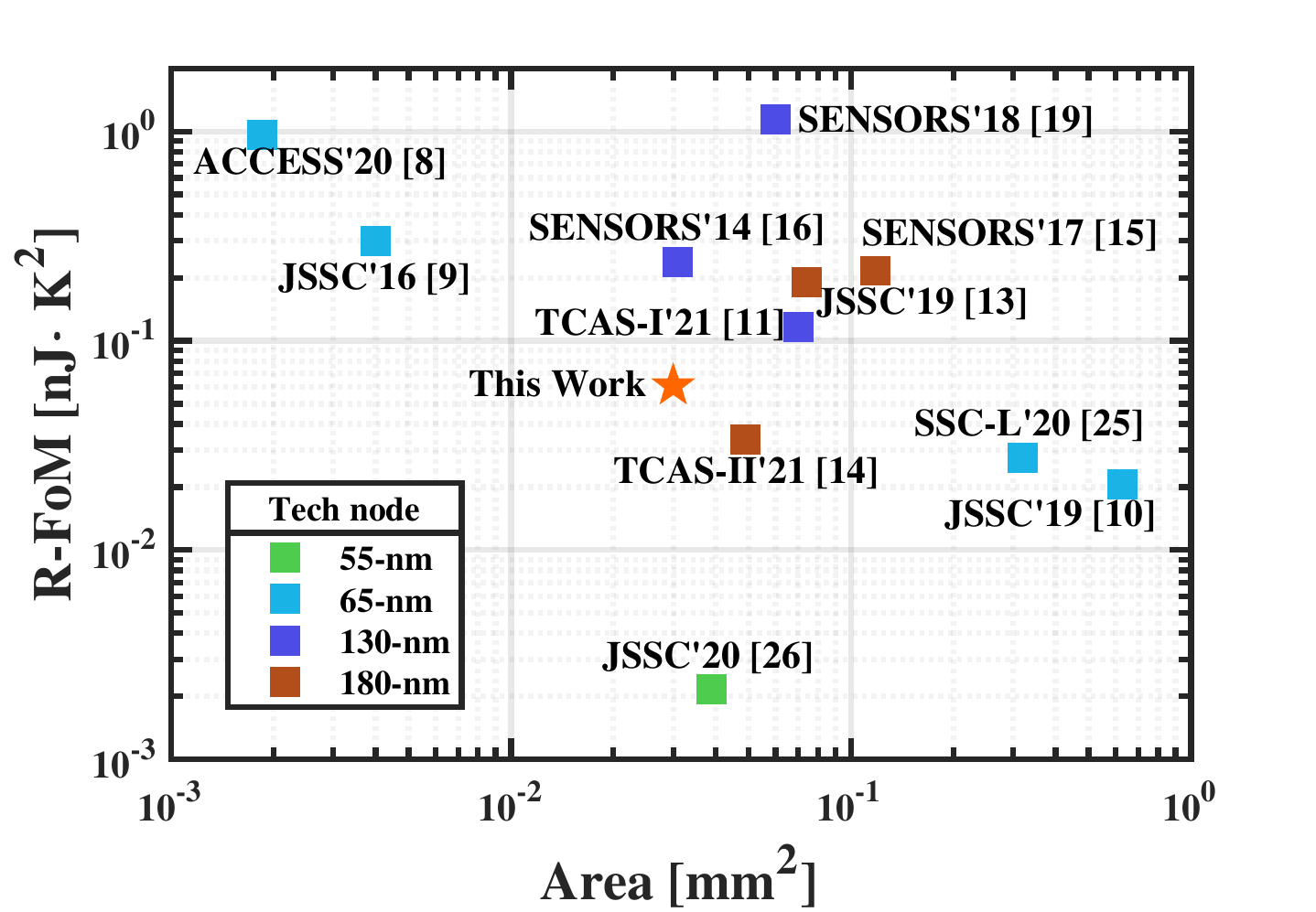}
     \caption{R-FOM vs silicon area.}
     \label{fig:FIG16}
     \vspace{-5mm}
   \end{figure}

After  a two-point calibration (i.e., with 10\C and 90\C as temperature calibration references), our sensor has a relative and a peak inaccuracy of 2.85\%  and of about 1.4\C, respectively (both well within the required specs for multi-core systems~\cite{vinshtok2020ultra,li2009}) in the 0\C $-$ 100\C temperature range. 
Such results are achieved without the need for additional non linearity correction logic, while exhibiting a noise limited resolution of \mbox{0.24\C} and an energy per conversion of only 1.06\,nJ. Above results lead to a competitive resolution figure-of-merit (R-FoM), defined as $Energy/Conversion \times Resolution^2$~\cite{Makinwa}. This can be graphically appreciated in \figref{fig:FIG16}, which shows the R-FoM/area trade-off for the compared designs. Reported results put our sensor in a position, where none of the competitors perform better in both 

\section{Conclusions} \label{sec:conclusions}
In this paper, we propose a fully-integrated CMOS  temperature sensor suitable for dense thermal monitoring in advanced SoC. Its low-complexity circuit topology allows compact footprint and voltage scalability along with low-power consumption and high accuracy in a large temperature range. The sensor was fabricated in 180\nm CMOS standard technology and experimentally characterized demonstrating competitive performance with respect to the state-of-the-art.
%\vspace{-3mm}

\bibliographystyle{IEEEtran}
\bibliography{bibliography.bib}
\vspace{-10mm}

\begin{IEEEbiography}[{\includegraphics[width=1in,height=1.25in,clip,keepaspectratio]{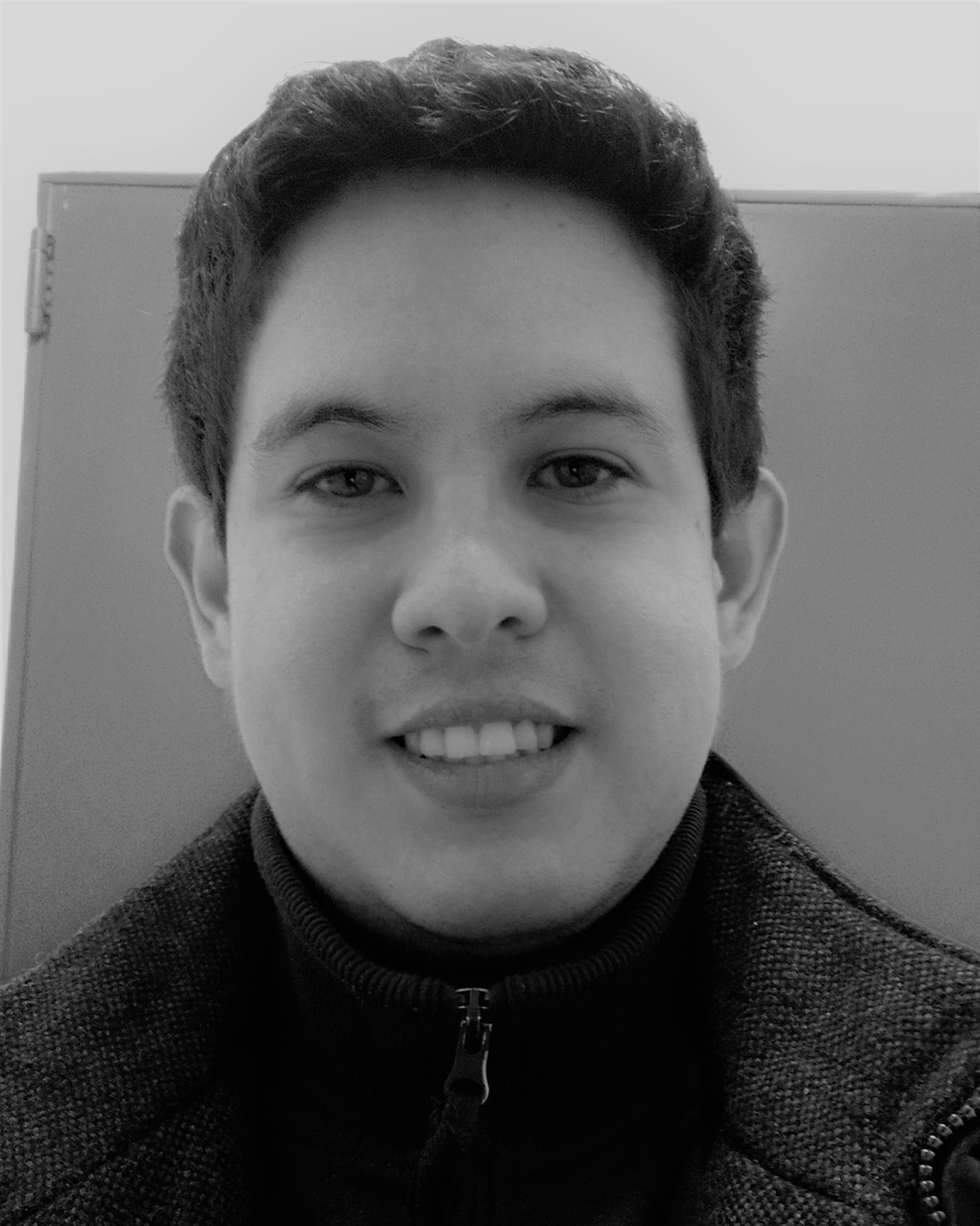}}]{Benjamin~Zambrano}
received the B.E degree in electronics from Escuela Superior Politecnica del Litoral, Guayaquil, Ecuador, in 2015, and a double M.S degree in nanoelectronics and electronics from University San Francisco de Quito, Ecuador, and University of Calabria, Rende, Italy, respectively, in 2020.
He is currently pursuing the Ph.D degree in ICT with the University of Calabria.
His research interests include integrated temperature sensors, ultralow-power/voltage designs and analog-based machine learning circuits.
\end{IEEEbiography}

\begin{IEEEbiography}[{\includegraphics[width=1in,height=1.25in,clip,keepaspectratio]{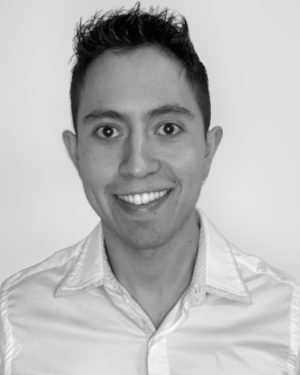}}]{Esteban~Garzón}
(S'14) received his B.Sc. degree in electronics engineering from Univ. San Francisco de Quito (USFQ), Ecuador, in 2016, and a double M.Sc. degree in nanoelectronics and electronics from USFQ and Univ. of Calabria (UNICAL), Italy, respectively, in 2018.
He is currently pursuing his Ph.D degree in the department of Computer Engineering, Modeling, Electronics and Systems Engineering, UNICAL, Italy.     
He has authored/co-authored more than 25 scientific papers in international journals and conferences.  
His research interests cover spin electronics/spintronics, cryogenic electronics, and emerging technologies for logic and memory low-power applications. 
\end{IEEEbiography}

\begin{IEEEbiography}[{\includegraphics[width=1in,height=1.25in,clip,keepaspectratio]{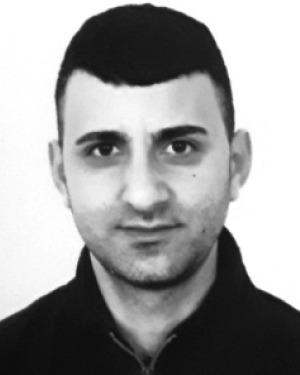}}]{Sebastiano~Strangio}
(Member, IEEE) was with the University of Udine, Udine, Italy, as a Temporary Research Associate, from 2013 to 2016, and with Forschungszentrum Jülich, Jülich, Germany, as a Visiting Researcher in 2015. From 2016 to 2019, he was with LFoundry, Avezzano, Italy. He is a Researcher in electronics at the University of Pisa, Pisa, Italy. He has authored and coauthored over 30 articles. His research interests include technologies for innovative devices (e.g., TFETs) and circuits for innovative applications [CMOS analog building blocks for deep neural networks (DNNs)], as well as CMOS image sensors, power devices, and circuits based on wide-bandgap materials. 
\end{IEEEbiography}

\begin{IEEEbiography}[{\includegraphics[width=1in,height=1.25in,clip,keepaspectratio]{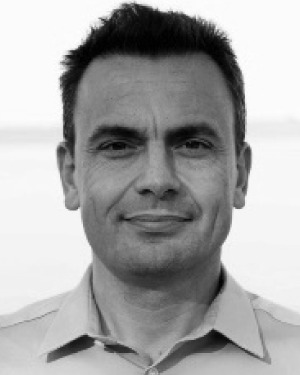}}]{Giuseppe~Iannaccone} (Fellow, IEEE) received the M.S. and Ph.D. degrees in electronic engineering from the University of Pisa, Pisa, Italy, in 1992 and 1996, respectively. He is a Professor of electronics at the University of Pisa. He has authored and coauthored more than 230 articles published in peer-reviewed journals and more than 160 papers in proceedings of international conferences, gathering more than 8500 citations on the Scopus database. His interests include quantum transport and noise in nanoelectronic and mesoscopic devices, development of device modeling tools, new device concepts and circuits beyond CMOS technology for artificial intelligence, cybersecurity, implantable biomedical sensors, and the Internet of Things, Dr. Iannaccone is a fellow of the American Physical Society.
\end{IEEEbiography}

\begin{IEEEbiography}[{\includegraphics[width=1in,height=1.25in,clip,keepaspectratio]{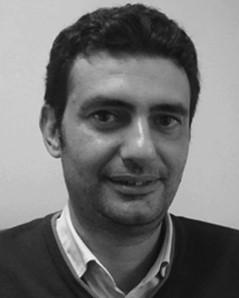}}]{Marco~Lanuzza}
(M’08–SM’16) received the Ph.D. degree in electronic engineering from the Mediterranea University of Reggio Calabria, Reggio Calabria, Italy, in 2005. Since 2006, he has been with the University of Calabria, Rende, Italy, where he is currently an Associate Professor. He has authored and coauthored more than 100 publications in international journals and conference proceedings. His recent research interests include the design of ultralow voltage circuits and systems, the development of efficient models and methodologies for variability-aware designs, and the design of digital and analog circuits in emerging technologies. Prof. Lanuzza is  an Associate Editor of Integration, the VLSI Journal.
\end{IEEEbiography}

\end{document}